\documentstyle[preprint,epsfig]{aastex}
\newcommand{\br}{\mbox{\it{\bf r}}}
\newcommand{\ba}{\mbox{\it{\bf a}}}
\newcommand{\bn}{{\mathbf{\nabla}}}

\shorttitle{Conservation Laws in SPH: the DEVA Code}
\shortauthors{Serna,  Dom\'{\i}nguez-Tenreiro, and S\'aiz}
\begin{document}

\title{
 Conservation Laws in Smooth Particle Hydrodynamics: the DEVA Code}
\author{ A. Serna$^{1}$, R. Dom\'{\i}nguez-Tenreiro$^{2}$, A.
S\'aiz$^{2}$}

\affil{$^{1}$Dept. de F\'{\i}sica y A.C., Universidad Miguel Hern\'andez,
03206 Elche, Alicante, Spain\\
 $^{2}$Dept. de F\'{\i}sica Te\'orica, Universidad Aut\'onoma de Madrid,
28040 Cantoblanco, Madrid, Spain}

\begin{abstract}
We describe DEVA, a multistep AP3M-like-SPH code particularly
designed to study galaxy formation and evolution in connection
with the global cosmological model. This code uses a formulation
of SPH equations which ensures both energy and entropy
conservation by including the so-called $\bn h$ terms. Particular
attention has also been paid to angular momentum conservation and
to the accuracy of our code. We find that, in order to avoid
unphysical solutions, our code requires that cooling processes
must be implemented in a non-multistep way.

We detail various cosmological simulations which have been
performed to test our code and also to study  the influence of the
$\bn h$ terms. Our results indicate that such correction terms
have a non-negligible effect on some cosmological simulations,
especially on high density regions associated either to shock
fronts or central cores of collapsed objects. Moreover, they
suggest that codes paying a particular attention to the
implementation of conservation laws of physics at the scales of
interest, can attain good accuracy levels in conservation laws
with limited computational resources.
\end{abstract}


\section{Introduction}
\label{introduction} In the last few years most cosmological
parameters have been determined up to a few percent. The values of
$\Omega_{\rm tot}$, $\Omega_{\Lambda}$, $\Omega_{\rm baryon}$ and
$H_0$  can now be constrained with an unprecedent degree of
accuracy \citep[see, for example,][and references
therein]{Lahav:02a,Lahav:02b,Nette:02,Sperg:03}. The next
challenge to cosmologists  is to test the
predictions of cosmological models at a few hundred kpc scales. It
turns out that these are just the relevant scales involved in
galaxy formation and evolution. Galaxy formation and evolution are
intriguing open questions whose resolution in connection with the
global cosmological model will very likely advance considerably in
this decade. Even though the field is at its beginnings, the use
of numerical methods to study how galaxies are assembled within a
cosmological scenario from the field of primordial fluctuations,
seems a convenient approach. The main advantage of these
approaches (i.e., self-consistent hydrodynamical simulations), is
that the physics is introduced at a very general level, and the
system evolves as a consequence. We can follow the evolution of
the dynamical and hydrodynamical properties of matter in the
Universe; galaxy-like objects (GLOs) appear as a consequence of
this evolution. And, so, the building-up of objects (cosmic
network structure formation at high $z$, collapse, interactions,
mergers, accretions), as well as their hydrodynamical consequences
(instabilities, gas infall from halos to discs at hundred of kpc
scales, gas inflow  along discs at tens of kpc scales, shocks,
cooling, piling-up of gas necessary for star formation), can be
accurately followed. We get not only the {\it properties\/} of
objects at any $z$, but also an {\it insight\/} into the {\it
physical processes\/} responsible for their formation and
evolution. Moreover, numerical hydrodynamical simulations using
particles permit very convenient comparisons of GLOs that form in
simulations with observational data. Simulations {\it directly\/}
provide us, at each $z$, with the structural and dynamical
properties of each individual GLO (position and velocity of each
of its  particles, gas density and temperature of each of its
baryonic constituents) and with their individual star formation
rates histories (SFRHs). Chemical abundance and
spectrophotometrical data are the current standard to compare
models of galaxy formation. It is expected that the next
generation of astronomical facilities will make possible a new
science: mass measurements for distant galaxies \citep[see, for
example][]{Verhe:02}. GLOs formed in numerical simulations are
particularly suited to be compared to this new kind of data.

Pre-prepared numerical experiments are adequate to describe in
detail a particular phase of the formation or evolution of
galaxies (for example, merger events or orbital motions of
satellites within halos), from initial conditions set by the
experimenter. These initial conditions try to model conditions
that would have arisen along the evolution of the systems under
consideration. They are useful to study basic aspects of the
physical processes relevant to evolution. For example, the works
by Barnes, Hernquist and Mihos
\citep{Barne:89,Barne:91,Barne:92,BarneH:92,Mihos:94,Mihos:96},
which have fundamental importance to understand the role played by
mergers in galaxy evolution, have been carried out with this
technique. However, in pre-prepared simulations, contrary to the
self-consistent approach, the process under consideration is
studied {\it in isolation}, and not in connection with the other
relevant processes involved in galaxy formation and evolution
(already mentioned) that, moreover, could interact among
themselves in a non-trivial way.

These considerations stress the ability of self-consistent
hydrodynamical simulations as a tool to learn how galaxies form
from the field of primordial fluctuations and evolve into the
objects we observe today.
To properly handle this problem, a numerical
code  has to
 allow for enough mass,
time and space resolution as well as a convenient dynamical range.
They should be as fast as possible and with memory requirements
within the current computer capabilities. A very important issue
when designing a numerical code to study galaxy formation and
evolution, is to make sure that conservation laws are accurately
verified, and, particularly, i), that angular momentum is
conserved at the scales relevant to disc formation; otherwise,
galaxy disc formation could meet with some difficulties; and, ii),
that entropy is conserved in reversible adiabatic processes,
because the violation of this physical principle could produce
spurious effects at galaxy scales. By the moment, star formation
(SF)  processes have to be modelled, either inspired in kpc or pc
scale hydrodynamical simulations
\citep{Padoa:99,Avila:01,Vazqu:00,Wada:01,Krits:02} or other
considerations
\citep{Katz:92,Tisse:97,Kenni:98,Yepes:97,Silk:01,Sprin:02b,Elmeg:03,Padoa:03}.

The first choice to be made when designing this kind of codes is
the gravity solver. Among current numerical methods, those that
employ adaptive techniques in regions of high density,  either
from a Lagrangian \citep[as AP3M, ][]{Couch:91,Couch:95} or
Eulerian \citep[as the
 ART and MLAPM codes,][]{Kravt:97,Knebe:01} approach, are the most suitable
to meet the requirements of resolution and large dynamical range,
accuracy and rapidity. A detailed comparison between AP3M and ART
codes has been carried out by \citet{Knebe:00}. They have found
out that these codes produce results that are consistent within a
10\%, provided that $ N_{\rm steps}/ DR \ge 2$ ($N_{\rm steps}$ is
the number of integration steps; $DR$ stands for the dynamical
range). The choice of the gravity solver in a cosmological
simulation depends on its purpose. To study galaxy formation and
evolution, Lagrangian codes  have the advantage over Eulerian
codes that they permit to go backwards and forwards in time in a
very easy way. For example, the constituent particles of a given
object can be identified at a given redshift, $z_1$, and one can
then analyze their positions in phase space and the properties of
the objects or structures they form at a different redshift,
$z_2$. This is a very convenient method  to study evolutionary
processes and it motivates our choice of an AP3M-based method as
gravity solver for our simulations.

To solve the hydrodynamical equations (and, in general, any
hyperbolic system of equations in partial derivatives), there are
also two basic different techniques: i), Eulerian methods, and,
ii), Lagrangian methods. Eulerian methods are based on the
so-called Godunov algorithm \citep{Godun:59}. Their new
formulations, using adaptive mesh refinements
\citep{Norma:98,Klein:98,Tey:02}, are particularly well suited to combine
with ART-like codes when both gravitational and hydrodynamical
forces are considered. For a comparison of the performances of a
number of hydrodynamical codes of both kinds see
\citep{Kang:94,Frenk:99}.

Most of the lagrangian methods used in astrophysics are based on
the SPH (Smooth Particle Hydrodynamics) technique
\citep{Lucy:77,Gingo:77,Monag:92}. Given the convenience of this
technique to be applied to cosmological simulations, a number of
authors developed SPH codes to be used in a cosmological context.
Some of them follow. \citet{Evrar:88} first used SPH techniques in
cosmological simulations (a P3M-SPH code).
\citet{Hernq:89,Katz:96}, as well as \citet{Navar:93}, coupled a
SPH code to the \citet{Barne:86} Tree algorithm. \citet{Vedel:99}
modelled their TREESPH code after \citet{Hernq:89} and
\citet{Dave:97} introduced a parallel version of this code, while
\citet{Stein:96} makes use of a special purpose hardware to
compute the gravitational forces by direct summation (GRAPE).
\citet{Serna:96} coupled SPH with a PP algorithm in a code
designed to be run on a Connection Machine, and \citet{Alimi:02}
incorporated in a Tree-SPH code the so-called $\bn h$ terms (see
below). GADGET \citep{Sprin:01} uses either a Tree scheme or
GRAPE, with individual integration timesteps, and both, serial and
parallel versions. Another parallel Tree-Sph code is GASOLINE
\citep{Borga:02}.

As already mentioned,  AP3M-based codes are particularly well
suited to study galaxy formation through self-consistent
cosmological simulations. The first AP3M-SPH code was introduced
by \citet{Couch:95} \citep[Hydra code, see also][]{Pearc:97}.
\citet{Tisse:97} carried out a second implementation. In these
implementations, the integration timestep is {\it global} (i.e.,
at a given time, the same for all particles). In cosmological
simulations, however, multiple time scales appear, due to their
very large dynamical ranges from very dense volumes to very
rarefied zones. To get an accurate enough integration scheme, and,
at the same time, to avoid that particles in denser volumes slow
down  the simulations, it is advantageous to introduce {\it
individual } integration timesteps, i.e., at each time, different
timesteps for each particle, depending on the density of the
region it samples. This is the optimal  design of the code
to increase the
mass resolution.

Another shortcoming  of conventional SPH formulations concerns the
entropy violation of the dynamical equations, related to the space
dependence of the smoothing length of  SPH particles, $h(r,t)$, as
noted by  some authors
\citep{Hernq:93,Nelso:93,Nelso:94,Serna:96}. A rigorous
formulation of SPH requires that additional terms must be included
in the particle equations of motion  which account for the
variability of $h$,  usually termed as ``the $\bn h$ terms''.
Until very recently, they were considered as having a negligible
effect on the global dynamics of systems \citep{Gingo:82,Evrar:88}
and, therefore, SPH codes ignored such additional terms and
focused on energy conservation. Alternatively, treatments of
hydrodynamics based on the Lagrange equations can be formulated
that are well behaved in their conservation properties of both,
energy and entropy, as that introduced recently by
\citet{Sprin:02a}\footnote{This paper is a reformulation of the
so-called {\it entropy} formulation of SPH, where dynamical
equations for the entropy instead of the energy were used
\citep{Lucy:77,BenzH:87,Hernq:93}, and where the energy
conservation was not guaranteed}. The effects of entropy violation
in SPH codes are not  completely clear and they need to be
analyzed in much more detail, specially in simulations where
galaxies are formed in a cosmological framework. Previous works
have analyzed this question in the case of the collapse of
isolated objects and have found that, if such correction terms are
neglected, the density peaks associated to central cores or shock
fronts are overestimated at a $\simeq$30$\%$ level
\citep{Alimi:02}.

To make up for these shortcomings when dealing with problems
related to galaxy formation and evolution, we introduce a new
code, DEVA, where gravity is solved by means of an AP3M-like
technique, and hydrodynamics with a SPH technique, with individual
integration timesteps. The space dependence of the SPH resolution
scale, $h(r,t)$, has been taken into account, in order to ensure
the conservative character of the equations of motion, as long as
entropy and energy is considered. Another  important particularity
of DEVA is the attention paid  to angular momentum conservation, a
key point to enable disc formation in simulations \citep[][ and
references therein]{Domin:98,Saiz:01}. Our choice has been to put
the stress into conservation laws rather than  into saving CPU time.
But saving CPU time has also been one of our concerns, so that the
code is fast enough  that cosmological self-consistent simulations
can be run on a modest computer machine.

The paper is organized as follows: $\S$1 is the Introduction. In
$\S$2, the SPH method is briefly reviewed and we present the SPH
equations when the $\bn h$ terms are considered. $\S$ 3 is devoted
to the particularities of the DEVA code, and $\S$ 4 to test
whether the code integrates correctly the hydrodynamical and
N-body equations (standard tests). In $\S$ 5 we introduce some
self-consistent simulations run with DEVA, compare  to one
standard of reference for hydrodynamical simulations in a
cosmological framework \citep[the Santa Barbara cluster comparison
project][]{Frenk:99}, and analyze the effects of the $\bn h$ terms
in these simulations. Finally, in $\S$ 6, we give a summary of the
work and discuss DEVA performances.

\section{The SPH Method}
\subsection{Kernel estimates}

The basic idea of the SPH method \citep{Lucy:77,Gingo:77} lies
in  representing the fluid elements by $N_g$ particles which act as
interpolation centers to determine the local value of any
macroscopic variable $f(\br)$. In order to smooth out local
statistical fluctuations, this interpolation is performed by
convolving the field $f(\br)$ with a smoothing (or kernel)
function $W$. For example, the smoothed estimate of the local
density is
\begin{equation}\label{rho}
\rho(\br_{i}) = \sum_{j=1}^{N_g} m_{j} W(r_{ij},h_{i},h_j)\;,
\end{equation}
where $ r_{ij}=\mid \br_i-\br_j\mid $, $m_j$ is the mass of
particle $j$, and $h_k$ is the smoothing length for particle $k$,
which specifies the size of the averaging volume.

Ideally, the individual particle smoothing lengths $h_k$ must be
updated so that each particle has a constant number of neighbors
$N_S$. By neighbors we mean those particles $j$ with distances
$r_{kj} \leq 2h_k$. Such a condition can be exactly implemented by
constructing, for each particle $k$, a list of its $N_S$ nearest
neighbors. The smoothing length of $k$ is then defined to be
\begin{equation}\label{hk}
h_k=\frac{1}{2}\mid \br_k-\br_{k_{f}}\mid\;,
\end{equation}
where $\br_{k_{f}}$ is the position vector of particle $k$'s most
distant neighbor. Since each particle has its own $h$ value, it is
possible to find couples of particles $(j,k)$ such that $j$ is a
neighbor of $k$, but $k$ is not a neighbor of $j$. In these cases,
it is obvious that the reciprocity principle\footnote{
The reciprocity principle states that, if at a given time
the $j$th particle belongs to the neighbor list of the $i$th
particle, then it is mandatory that, at this same time, the $i$th
particle belongs to the neighbor list of the $j$th particle}
 is not satisfied and,
therefore, simulations will not conserve momentum. In order to
solve this problem, it is necessary to symmetrize the SPH
equations by using, for example, averaged kernels
\citep{Hernq:89}:
\begin{equation}\label{Wij}
W_{ij}\equiv W(r_{ij}, h_i, h_j)=\frac{1}{2}\left[ W(r_{ij},h_i)+
W(r_{ij},h_j)\right]\;.
\end{equation}

A first consequence of the adopted symmetrization procedure is the
specific form for the kernel derivatives. As a matter of fact, Eq.
(\ref{Wij}) implies that $W_{ij}$ is a function of three
variables: $r_{ij}$, $h_i$ and $h_j$. Consequently, its gradient
$\bn_i W_{jk}$ is given by:
\begin{eqnarray}\label{nablaWij}
 \bn_i W_{jk} = \frac{1}{2} \left[\left(\frac{\partial W(r_{jk},h_j)}{\partial r_{jk}} +
\frac{\partial W(r_{jk},h_k)}{\partial r_{jk}}\right) \bn_i r_{jk}\right] \nonumber \\
 + \frac{1}{2}\left[\frac{\partial W(r_{jk},h_j)}{\partial h_{j}}\bn_i h_{j} +
\frac{\partial W(r_{jk},h_k)}{\partial h_{k}} \bn_i h_{k}\right]
\end{eqnarray}
The first part of Eq. (\ref{nablaWij}), which does not involve
derivatives of the smoothing lengths, is the usual symmetrized
form of $\bn_i W_{jk}$. The second part, which involves
derivatives of the smoothing lengths, arises because of the
spatial and temporal variability of $h$. We shall refer to terms
of this type as '$\bn h$ terms'. Most implementations of the SPH
algorithm consider only the first one and neglect the $\bn h$
terms.

\subsection{Hydrodynamic equations}

The motion of particle $i$ is determined by the momentum and
energy equations:
\begin{eqnarray}
\frac{d\mbox{\it{\bf v}}_{i}}{dt}  &=&
\textbf{a}_{i}^{P}+\textbf{a}_{i}^{visc}
 - \bn\Phi_{i} \label{accel}\\
 \frac{du_i}{dt} &=& \frac{P_i}{\rho_{i}^{2}}\frac{d\rho_i}{dt}+{\cal{H}}_i\;,
 \label{dedt}
\end{eqnarray}
where $\Phi_{i}$ is the local gravitational potential,
$\textbf{a}_{i}^{P}$ is the acceleration due to pressure forces,
$\textbf{a}_{i}^{visc}$ is the acceleration due to viscosity
forces, $u$ is the specific internal energy,
 $P=(\gamma -1)\rho u$ is the pressure (with $\gamma$ being
the constant heat ratio), and ${\cal{H}}_i$ is the power due to
non-adiabatic heating or cooling processes.

A fully consistent SPH expression for pressure forces, satisfying
all conservations laws (including entropy conservation in
reversible adiabatic problems), was obtained by
\citet{Nelso:93,Nelso:94}:
\begin{equation}\label{Pforce} {\bf F}_{i}^{P}=-\sum_j
m_j\frac{P_j}{\rho_{i}^{2}}\bn_i\rho_j\;.
\end{equation}

Using Eqs. (\ref{rho}) and (\ref{nablaWij}) to compute
$\bn_i\rho_j$, one obtains
\begin{eqnarray}\label{accpress}
{\bf a}_{i}^{P}&=& -\sum_j m_j\left(\frac{P_i}{\rho_{i}^{2}}+
\frac{P_j}{\rho_{j}^{2}}\right) \left[\frac{\partial
W(r_{ij},h_i,h_f)}{\partial r_{ij}} \right]
\frac{\br_{ij}}{r_{ij}}
\nonumber\\
&-&\frac{1}{4}\tilde{r}_i \sum_j
m_j\left(\frac{P_i}{\rho_{i}^{2}}+ \frac{P_j}{\rho_{j}^{2}}\right)
\frac{\partial W(r_{ij},h_i)}{\partial h_{i}}\\
&+&\frac{1}{4}\sum_j \tilde{r_j}\delta_{ij_f} \sum_k \frac{m_j
m_k}{m_i} \left(\frac{P_j}{\rho_{j}^{2}}+
\frac{P_k}{\rho_{k}^{2}}\right) \frac{\partial
W(r_{jk},h_j)}{\partial h_{j}}\;, \nonumber
\end{eqnarray}
where $\tilde{r}_k\equiv\br_{kk_f}/r_{kk_f}$.

On the other hand, using Eqs. (\ref{rho}) and (\ref{nablaWij}) to
compute the $d\rho_i/dt$ derivative appearing in Eq. (\ref{dedt}),
the energy equation becomes:
\begin{eqnarray}\label{energy}
 \frac{du_i}{dt}&=&\frac{P_i}{\rho_{i}^{2}}\sum_j m_k \left[
 \frac{\partial W(r_{ij},h_i,h_j)}{\partial r_{ij}}\right]
 \frac{\textbf{r}_{ij}\cdot\textbf{v}_{ij}}{r_{ij}}\nonumber\\
 &+& \frac{1}{4}\frac{P_i}{\rho_{i}^{2}}\sum_j m_j
 \left[\frac{\partial W(r_{ij},h_i)}{\partial h_i}
 \breve{r}_i+
\frac{\partial W(r_{ij},h_j)}{\partial h_j}
 \breve{r}_j\right] + {\cal{H}}_i\;,
\end{eqnarray}
where
\begin{equation}
\breve{r}_k\equiv\frac{\textbf{r}_{kk_f}\cdot\textbf{v}_{kk_f}}{r_{kk_f}}\;.
\end{equation}

Note that Eqs. (\ref{accpress}) and (\ref{energy}) have been
deduced by using
 both spatial and time derivatives of the SPH
density as defined by Eq. (\ref{rho})
{\it with the symmetrization specified in Eq. (3)},
because
compatibility  with the  conservation laws  requires that the
SPH force and energy equations  are evaluated in consistency with
the density definition.
In the case of DEVA, this requirement increases the CPU time
per integration step.
In fact, since the density $\rho_i$ associated to a particle
$i$ depends on both $h_i$ and $h_j$, for $j = 1, ..., N_S$,
(i.e., for its $N_S$ nearest neighbors, see Eq. \ref{rho}), the
computation of $\rho_i$  at a given  integration step requires the
knowledge of  $h_j$ for these $N_S$ nearest neighbors at its beginning
\footnote{ As a matter of fact,
each $h_j$ value must be  kept fixed all  along   the
integration step in order to
avoid violating the reciprocity principle}.
 This can be achieved
either by using the $h_j$ values predicted in the previous
integration step or by performing, at each step, a first loop over
the particles to compute their $h_j$ values and, once it is
over, a second loop to compute their hydrodynamical
properties. Since we look for a high accuracy rather than a high
computational speed, we have adopted this latter possibility.

As usual in SPH, to account for dissipation at shocks, the above
equations must be completed by adding an artificial viscous
pressure term, $\Pi_{ij}$. When the $\bn h$ terms are considered,
$\Pi_{ij}$ is added   only to  the leading term of equations
(\ref{accpress}) and (\ref{energy}),  that is, those not involving
$\bn h$ terms \citep{Nelso:94}:
\begin{equation}
\frac{P_i}{\rho_{i}^{2}}\rightarrow\frac{P_i}{\rho_{i}^{2}}+\frac{\Pi_{ij}}{2},
\qquad
\frac{P_i}{\rho_{i}^{2}}+\frac{P_j}{\rho_{j}^{2}}\rightarrow\Pi_{ij}\;,
\end{equation}
where we have adopted the standard viscous pressure proposed by
\citet{Monag:83}:
\begin{equation}
\Pi_{ij} = \frac{-\alpha\mu_{ij}\overline{c}_{ij}+
\beta\mu_{ij}^{2}}{\overline{\rho}_{ij}}\;,\label{Pi}
\end{equation}
where $\alpha$ and $\beta$ are constant parameters of order unity,
$\eta^2$ is a softening parameter to prevent numerical
divergences, $c_i$ is the local sound speed, and
\begin{equation}
\mu_{ij} = \left\{
\begin{array}{ll}
\frac{\mbox{\it{\bf v}}_{ij}\br_{ij}}{{\normalsize
h_{ij}\left(r^{2}_{ij}/h_{ij}^{2} +\eta^{2}\right)}}
& \mbox{\it{\bf v}}_{ij}\br_{ij}<0\\
0 & \mbox{\it{\bf v}}_{ij}\br_{ij}\ge 0
\end{array}\right.\;.
\end{equation}

\section{The DEVA code}

The simulation of a system constituted by $N$ particles usually
requires a computational effort which considerably varies from
some regions (or particles) to other. For example, regions of high
density and submitted to strong shocks need to be simulated with
timesteps much shorter than the rest of the system. In the
AP3M+SPH codes described in the literature, all the particles in
the system are simultaneously advanced at each timestep. The
particle needing the highest time resolution determines the
timestep length of all the others. Consequently, some few
particles can slow down the simulation of a system. To make a code
more efficient in handling  with problems with multiple time scales, the
computational effort must be centered on those particles that
require it, avoiding useless computations for the remaining
particles. In other words, it is necessary to allow for different
timesteps for each particle.

\subsection{AP3M with individual timesteps}

A PEC (Predict-Evaluate-Correct) scheme with individual timesteps
has been developed and implemented on our code in the following
way:
\begin{enumerate}
 \item We enter the step $n$ (which corresponds to the time $t^n$) with
known positions $\br_{i}^{n}$, velocities $\mbox{\it{\bf
v}}_{i}^{n}$, and accelerations $\ba_{i}^{n}$, for all the $N$
particles. Furthermore, any integration scheme with individual
timesteps needs some information to identify, at each step, those
particles needing a recomputation of their acceleration. This
information is stored in two vectors $t_{i}^{last}$ and
$t_{i}^{next}$, where $t_{i}^{last}$ is the time at which the last
update of $\ba_i$ was performed, while $t_{i}^{next}=t_{i}^{last}
+ \Delta t_i$ is the time at which a recomputation of $\ba_i$ will
be necessary in the future.
\begin{equation}\label{variables}
 \br_{i}^{n}, \mbox{\it{\bf v}}_{i}^{n}, \ba_{i}^{n}, t_{i}^{next}, t_{i}^{last}
\end{equation}
\item A list is constructed with those particles $j$ which will be
advanced at the current step. Such particles are labelled as {\it
active}. Obviously, the particle $j_{min}$ with the smallest
prediction time, $t_{j}^{next}-t^n$, must be included in this
list, and fixes the timestep of the remaining active particles:
 \begin{equation}\label{Deltat}
 \Delta t =\min_{j}(t_{j}^{next}-t^n)\;.
 \end{equation}
Since each step requires the update of many auxiliary arrays, it
is impractical to advance only a single particle. For this reason,
we label as \textit{active} all particles within a cubic box
around $j_{min}$. The size of the activation box is chosen, at
each position, so that it contains a small fraction of the total
number of particles.
\item For all particles, active or not, we predict the value of
$\br^{n+1}$ and $\mbox{\it{\bf v}}^{n+1}$ at $t^{n+1}$
 \begin{eqnarray}
 \tilde{\br}_{i}^{n+1} &=& \br_{i}^{n}+\mbox{\it{\bf v}}_{i}^{n}\Delta t+ \ba_{i}^{n}(\Delta t)^2/2\\
 \tilde{\mbox{\it{\bf v}}}_{i}^{n+1} &=& \mbox{\it{\bf v}}_{i}^{n}+\ba_{i}^{n}\Delta
 t\;.
 \end{eqnarray}
\item Only for active particles, we compute their accelerations
and correct $\br^{n+1}$ and $\mbox{\it{\bf v}}^{n+1}$ using :
 \begin{eqnarray}
 \br_{j}^{n+1} &=& \tilde{\br}_{i}^{n+1}+A(\ba_{j}^{n+1}-\ba_{j}^{n})
 (\delta t_j)^2/2\\
 \mbox{\it{\bf v}}_{j}^{n+1} &=& \tilde{\mbox{\it{\bf v}}}_{i}^{n+1}+B(\ba_{j}^{n+1}-\ba_{j}^{n})
 (\delta t_j)\;,
 \end{eqnarray}
where the choice $A=1/3$ and $B = 1/2$ maintains accuracy to
second order both in positions and velocities.
In these expressions, $\delta t_j$ represents the time interval
elapsed from the last evaluation of $\ba_j$ to that performed in
the current timestep
 \begin{equation}
 \delta t_j=t^n+\Delta t-t_{j}^{last}\;.
 \end{equation}
Note that, unlike $\Delta t$, the $\delta t_j$ value is different
for each active particle.
\item We update the global time $t^{n+1}=t^n+\Delta t$, as well as the
$t_{j}^{last}$ and $t_{j}^{next}$ values of each active particle.
Here, in order to maintain the numerical stability of the AP3M
algorithm, the individual timestep $\Delta t_j$ must be smaller
than the time scale for significant displacements or changes in
velocity due to accelerations:
\begin{equation}
\Delta t_{i}^{a} =(\epsilon^{2}/{a_{i}^2})^{1/4}\;,
\end{equation}
where $\epsilon$ is the gravitational softening.
\end{enumerate}

\subsection{Including SPH}\label{IncSPH}

The above integration scheme may easily be extended to include
hydrodynamics. The SPH processes involve three new independent
variables in addition to those listed in Eq. (\ref{variables}):
\begin{equation}
h_{j}^{n}, u_{j}^{n}, \dot{u}_{j}^{n}\;,
\end{equation}
where $h_{j}^{n}$ is the smoothing length, $u_{j}^{n}$ the
specific internal energy, and $\dot{u}_{j}^{n}$ its derivative.
For all particles, we must then predict the value of $u_{j}^{n+1}$
at $t^{n+1}$
\begin{equation}\label{epredict}
   \tilde{u}_{j}^{n+1} = u_{j}^{n}+\dot{u}_{j}^{n}\Delta t\;,
\end{equation}
and compute, for active particles, both their total acceleration
($\ba_{grav}^{n}$ and $\ba_{hydro}^{n}$) as well as their
hydrodynamical variables ($h^{n+1}$, $\rho^{n+1}$, $P^{n+1}$ and
$\dot{u}_{i}^{n}$). These quantities are then used to correct the
internal energy of active particles:
\begin{equation}
   u_{j}^{n+1} = \tilde{u}_{j}^{n+1}+C(\dot{u}_{j}^{n+1}-\dot{u}_{j}^{n})
 (\delta t_j)\;,
\end{equation}
where the choice $C=1/2$ maintains accuracy to second order in
internal energies.

Now, the numerical stability requires additional limits on the
timestep of each gas particle. A first timestep control is that
concerning the time scale for significant displacements or changes
in velocity due to accelerations:
\begin{equation}
\Delta t_{i}^{a} =(h_{i}^{2}/{a_{i}^2})^{1/4}\;.
\end{equation}
A second limit on $\Delta t_i$ is usually given by a timestep control which
combines the Courant and the viscous conditions:
\begin{equation}
\Delta t_{i}^{cv} = \left[ \frac{h_{i}}{c_{i}+1.2(\alpha c_{i} +
\beta \max_{j}\mid\!\mu_{ij}\!\mid)}\right]\;.\label{Dtcv}
\end{equation}

When required, radiative cooling is implemented in an integral
form \citep{Thoma:92} using the fact that, due to the Courant
condition, the density field is nearly constant over a time-step:
\begin{equation}
\int_{u_{i}}^{u_{i}-\Delta u_{i}^{cool}}
\frac{du_i^{cool}}{\Lambda} = -\frac{\Delta
t}{\rho_i}\label{ucoolint}\;,
\end{equation}
where $\Lambda (T,\rho)$ is the power radiated per unit volume and
$\Delta u_{i}^{cool}$ is the change in $u_i$ due to cooling
processes.

This integral procedure circumvents the need of a control time for
cooling, and, hence, it never limits the timestep. The numerical
stability of our code requires that cooling effects must be
updated at each step for all particles, active or not. Otherwise,
the simultaneous presence of already cooled and not yet cooled
particles in a given object would break the local pressure
equilibrium and, as a result, cold particles would fall to the
object center causing a non-physical core of very high density
(see $\S$5.2 for an example).

\begin{figure}
\centerline{\epsfig{figure=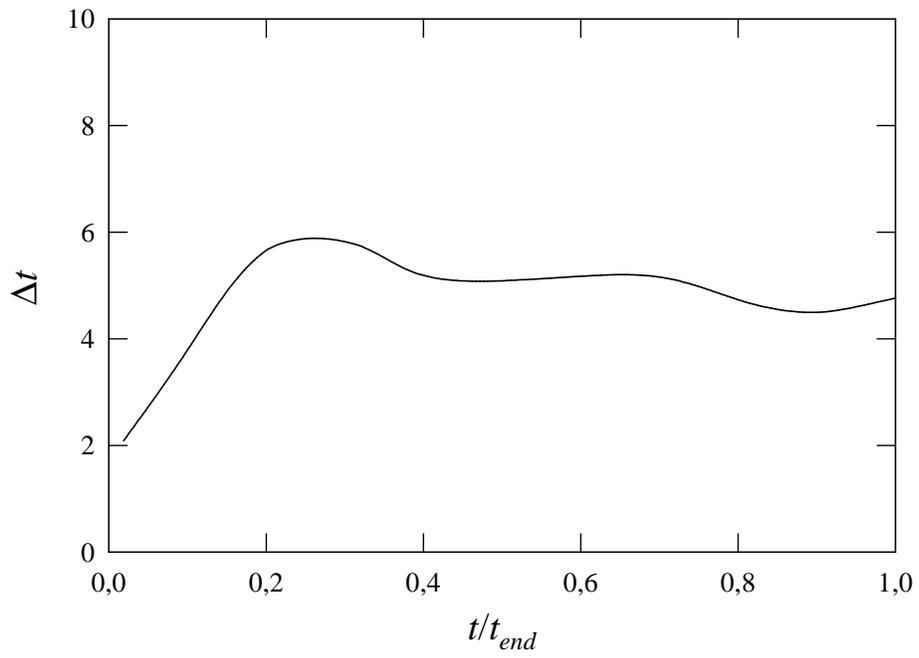,width=0.75\textwidth} }
\caption{Ratio of the CPU time consumed by an algorithm where all
particles are simultaneously advanced to that consumed when
individual timesteps are considered. } \label{CPU}
\end{figure}

Fig \ref{CPU} shows, for a typical cosmological simulation, the
ratio of the CPU time consumed by an algorithm with individual
timesteps to that consumed when all particles are simultaneously
advanced. We see that the use of individual timesteps typically
reduces the CPU time per step in a factor of five.
In a pentium IV 1.7GHz personal computer, the CPU time typically
consumed by our code in a $2\times64^3$ cosmological simulation
without the $\bn h$ terms is: a) 25 seconds per step in
simulations without radiative cooling (such as the Santa Barbara
cluster test of \S 5.1), b) from 25 (at high redshifts) to 70 (at
low redshifts) seconds per step in cosmological simulations with
radiative cooling (such as those of \S 5.2). These CPU times are
increased by about 150\% when the $\bn h$ terms are taken into
account.

\section{Adiabatic Tests}

DEVA has been applied to different problems with known analytical
or numerical solutions. The aim of such simulations was not only
to test our code, but also to analyze the effects of the $\bn h$
terms included in it.

\subsection{The one-dimensional shock tube problem}

\begin{figure}
\centerline{\epsfig{figure=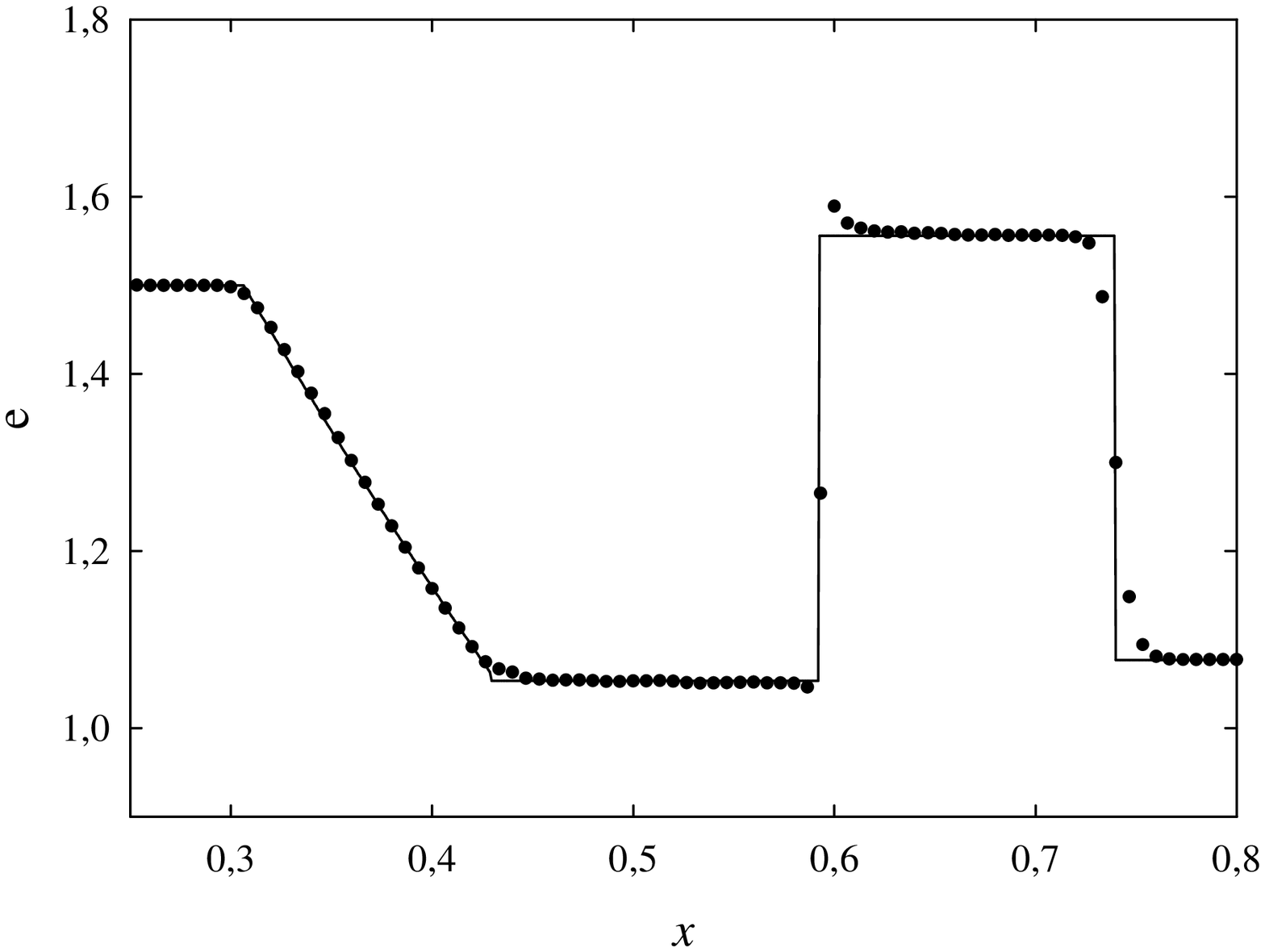,width=0.5\textwidth}
\epsfig{figure=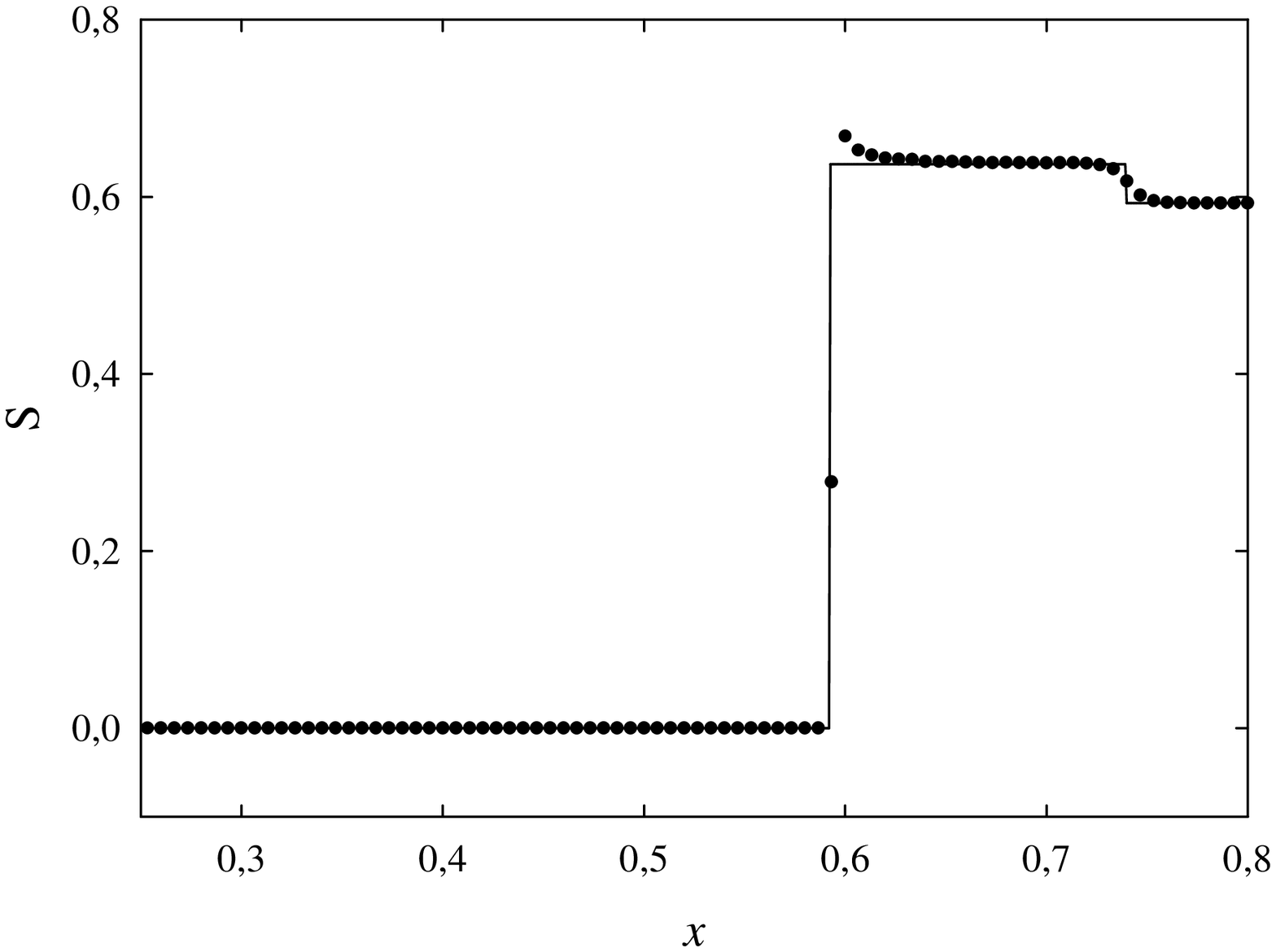,width=0.5\textwidth}}\vspace{0.5cm}

\centerline{\epsfig{figure=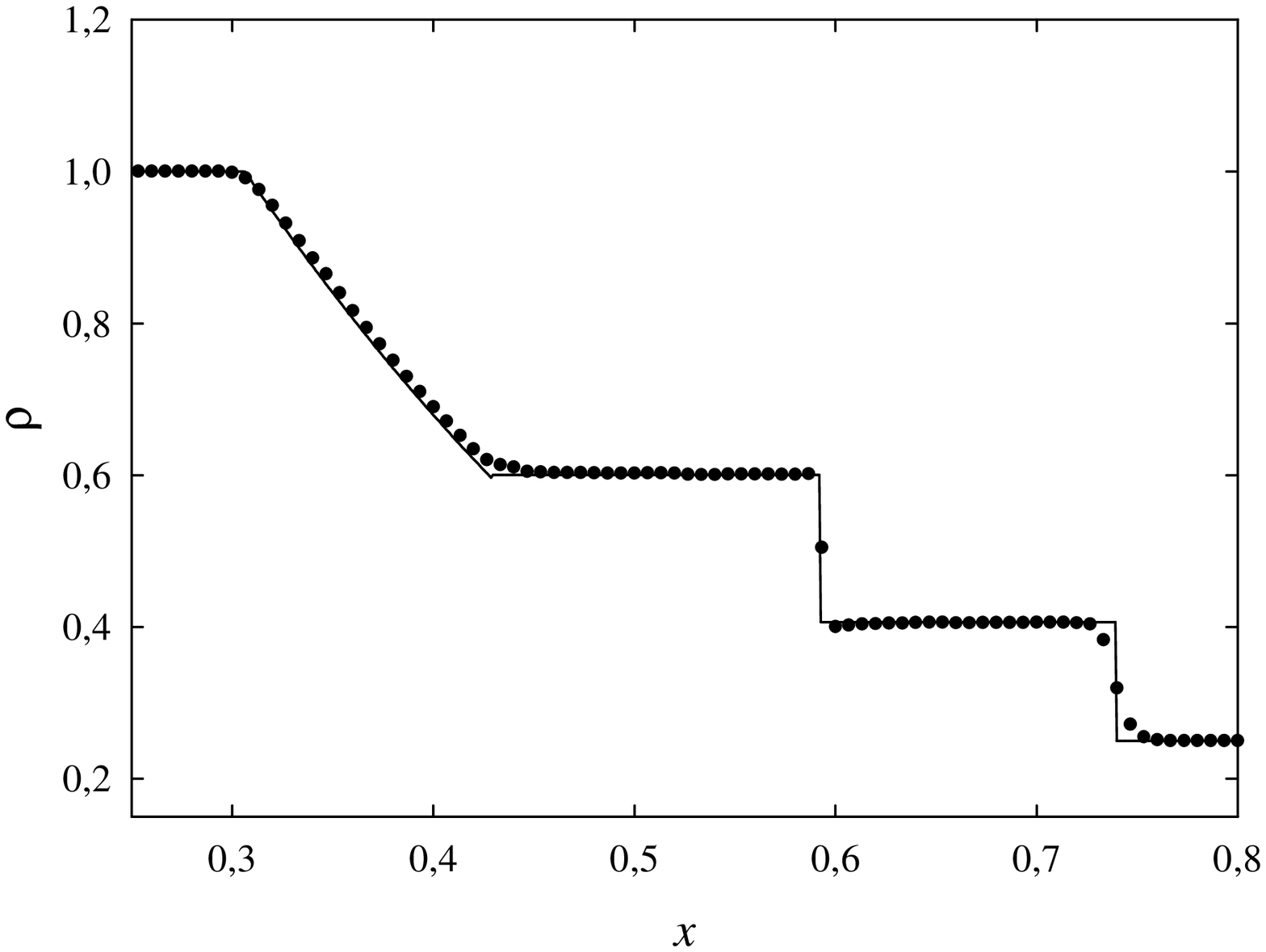,width=0.5\textwidth}
\epsfig{figure=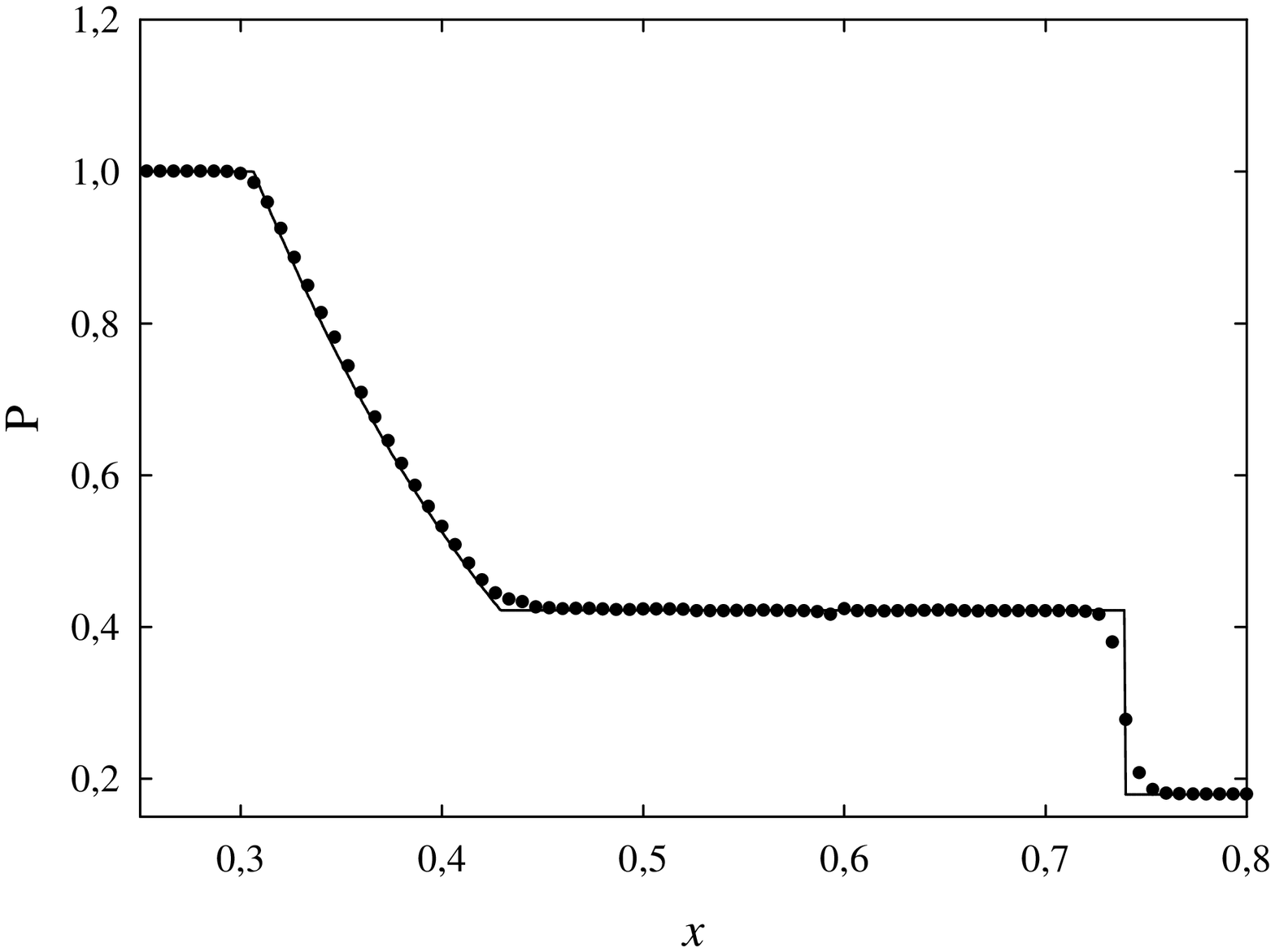,width=0.5\textwidth}}

\caption{a) Thermal energy, b) entropy, c) density, and d)
pressure profiles at $t=0.15$ in the one-dimensional shock tube
problem. Points represent the DEVA results, and solid lines are
the analytical solutions} \label{shocktube}
\end{figure}

The one-dimensional shock tube problem proposed by \citet{Sod:78}
has become a standard test of all transport and source terms
(including artificial viscosities) of hydrodynamic algorithms. It
considers a perfect gas distributed on the $x$-axis. A diaphragm
at $x_0$ initially separates two regions which have different
densities and pressures. All particles are initially at rest. At
time $t=0$ the diaphragm is broken and both regions start to
interact. Nonlinear waves are then generated at the discontinuity
and propagate into each region: a shock wave which moves from the
high to the small pressure region, while the associated
rarefaction wave moves in the inverse sense. The analytical
solution to this problem has been given by \citet{Hawle:84} and
\citet{Rasio:91}. In our simulation, we have considered $N=4096$
gas particles initially distributed in the interval $0\le x < 1$
according to:
\begin{eqnarray}
& \rho=1\;\; P=1\;\; v=0\;\;\; & (\mbox{for\hspace{1ex}} x<0.5)\nonumber\\
& \rho=0.25\;\; P=0.1795\;\; v=0\;\;\; & (\mbox{for\hspace{1ex}}
x\ge0.5) \nonumber
\end{eqnarray}
Dissipational effects, other than those associated with the
artificial viscosity (with $\alpha=\beta=1$ and $\eta^2=0.01$),
were ignored, as well as gravitational interactions. Fig.
\ref{shocktube} shows our results at $t=0.15$. We see from this
figure that our results are in excellent agreement with the
analytical solutions. The resulting profiles both in the shock
wave (located at $x\simeq 0.75$) and in the contact discontinuity
(located at $x\simeq0.6$) are much less rounded than in previous
SPH computations \citep[see, e.g.][]{Monag:83,Hernq:89,Rasio:91}
as a result of having used a larger number of particles and,
hence, a better resolution. We also note the almost complete
suppression of post-shock oscillations in our results. These
oscillations can be seen in the previous SPH simulations of this
problem, especially in the velocity field, while no high-frequency
vibrations are perceptible in our results. The weak blip observed
in the pressure profile at the contact discontinuity
($x\simeq0.6$) is normal in SPH codes. Such non-physical blip has
been explained by \citet{Monag:83} as due to the fact that the
smoothed estimate of pressure is computed by using discontinuous
quantities. It is then inevitable that $P$ has some slight
perturbation at the contact discontinuity, but it has a negligible
effect on the motion. In this test, simulations including the $\bn
h$ terms gave exactly the same results as those neglecting such
terms.

\subsection{Adiabatic collapse of a non-rotating gas sphere}\label{collapse}

\begin{figure}
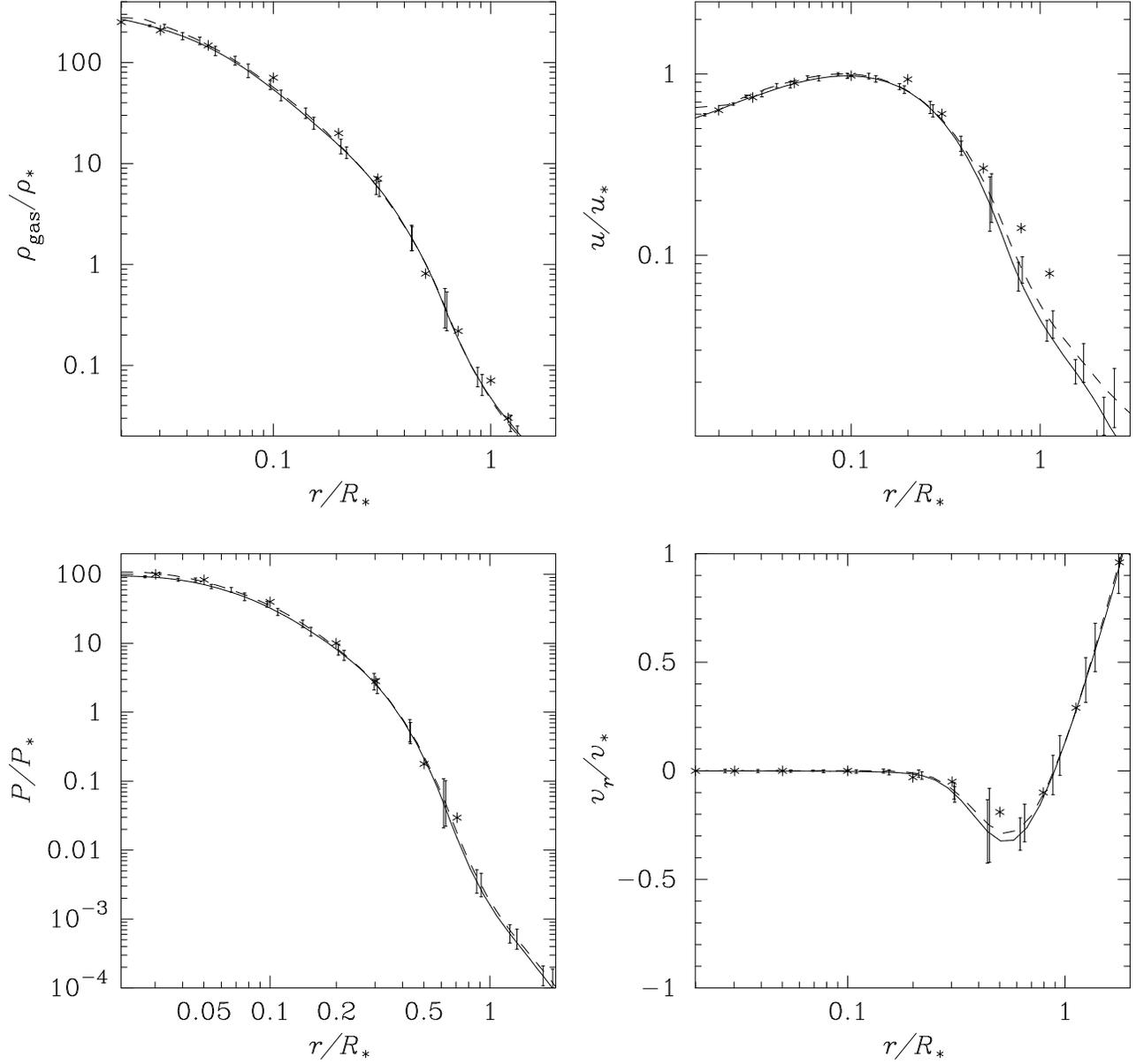

\centerline{\epsfig{figure=f3a.eps,width=0.5\textwidth}\hspace{0.25cm}
\epsfig{figure=f3b.eps,width=0.5\textwidth}}\vspace{0.5cm}

\centerline{\epsfig{figure=f3c.eps,width=0.5\textwidth}\hspace{0.25cm}
\epsfig{figure=f3d.eps,width=0.5\textwidth}}

\caption{ Final gas density, thermal energy, pressure and radial
velocity profiles for the adiabatic collapse of a non-rotating gas
sphere. Stars: \citet{Hernq:89} results. Solid (dashed) line: DEVA
results when the $\bn h$ terms are (are not) taken into account,
with their corresponding errors (see text) }\label{hkfig}
\end{figure}

A 3D-problem usually considered to  test hydrodynamical codes is
that concerning the adiabatic collapse of a non-rotating gas
sphere.  This problem has  been  studied from a finite-difference
method by \citet{Thoma:87}, and  from SPH simulations  by
\citet{Evrar:88} and \citet{Hernq:89}. In order to facilitate  the
comparison of our results to those obtained  by these authors,  we
have taken  their same initial conditions: a gas sphere of radius
$R$ and total mass $M_T$, with density profile
\begin{equation}
\rho = \frac{M_T}{2\pi R^2}\frac{1}{r}\;.
\end{equation}
All the  $N=4096$ gas  particles are initially at rest and  have
the same specific internal  energy $u = 0.05GM_{T}/R$. Units were
taken  so that $2G=M_T=2R=1$. Initially far from equilibrium, the
system  collapses converting most of  its   kinetic energy into
heat.   A slow expansion follows and, at  late times, a core-halo
structure develops with   nearly isothermal inner regions and the
outer  regions cooling adiabatically. We show in Fig. \ref{hkfig}
different system profiles at end of the simulation. The solid line
represents the numerical solution obtained when the $\bn h$ terms
have been included, the dashed line represents the numerical
solution obtained when these terms have been neglected, and the
points represent the numerical solution obtained by
\citet{Hernq:89}. We see that, although the solid and dashed lines
are not exactly superposed, both solutions are coincident within
the error bars.

\subsection{Interpretation of the influence of the $\bn h$ terms}

We can understand why the $\bn h$ terms have a negligible effect
in the two standard tests reported in this section. The effect of
the $\bn h$ terms on the thermal energy can be expressed by a time
scale defined by
\begin{equation}\label{ts}
  t_S=\langle t_{S}^{i}\rangle\;,
\end{equation}
where $t_{S}^{i}$ is the ratio of the specific thermal energy
$u_i$ of particle $i$ to the change in $u_i$ due to the $\bn h$
terms
\begin{equation}\label{tsi}
  t_{S}^{i}=\frac{u_i}{\dot{u}_{i}^{\bn h}}\;.
\end{equation}
When the time elapsed, $t$, is longer than this time scale, the
$\bn h$ terms will produce a non-negligible effect on the thermal
energy. This condition can be expressed as
\begin{equation}
  I=\int_{0}^{t}\frac{dt'}{t_S(t')}\ge 1\;,
\end{equation}
where $I$ represents the area contained by the $1/t_S(t')$ curve
between $t'=0$ and $t'=t$.

When these timescales are computed for the non-adiabatic tests
reported in this paper, one obtains $I=0.001$ (in the shock tube
problem), and $I=0.85$ (in the collapse of a non-rotating gas
sphere). The effect of the $\bn h$ terms are then expected to be
small.

\subsection{Quntifying the effects of the $\nabla h$ terms}

Testing the effects of the $\nabla h$ terms on hydrodynamical
evolution can be better worked out in isentropic processes. It is
not easy, however, to get such kind of processes in simulations of
gas evolution because gas easily develops shocks where dissipation
must occur. A possibility is considering the adiabatic
expansion of a gas sphere
from a situation of equilibrium. Using expansion rather
than collapse has the advantage that shell crossing decreases
substantially, so that viscous force terms can be removed and the
evolution is isentropic.

Such simulations  as   that   shown in   \S  \ref{collapse} lead at
late times, $t\gtrsim3$,  to   equilibrium   spheres with
density  profiles  as that displayed in Fig. \ref{hkfig}.
We use the adiabatic
expansion of such  spheres   to test the effects
of the $\bn h$ terms in DEVA. Initial conditions were generated by
performing a simulation as  that described in \S \ref{collapse},
using different numbers of particles. At  $t=3$, we switched-off its
viscous pressures (by setting $\alpha=\beta=0$)   to
ensure that the subsequent evolution  conserves the total
entropy. The self-gravity was also switched-off at $t=3$.
In absence of gravitational interactions, this system expands
fast  and,  at $t=3.3$, its central density has decreased  by a
factor  of $\approx 25$. The evolution from $t=3$ to  $t=3.3$ must
conserve both the total energy and entropy.

Table 1 shows the results for this series of simulations. We see
that, when  the $\bn h$ correction terms are neglected, energy is
conserved  very accurately but there  exists a considerable
violation  of the total  entropy (about $5\%$   in the
time interval we have considered).
 In the opposite case, when such correction terms
are taken  into account,  both total energy and total entropy are
conserved very accurately  (about $0.02\%$). These results appear
to be independent on the number of particles.
We then conclude that the $\bn h$ correction terms cure entropy
violation, allowing, at the same time, a very good energy
conservation.

\section{Self-Consistent Cosmological Simulations}

The DEVA code is particularly well suited to numerically follow,
in a cosmological context,
the assembly  from the field of primordial fluctuations
of collapsed objects, such as galaxy clusters  or
galaxies.
 To illustrate DEVA performances in this
situation, we briefly analyze some results of self-consistent
simulations. Self-consistency means that initial conditions are
set at high $z$ as a Montecarlo realization of the field of
primordial  fluctuations (i.e., perturbations,
characterized by a spectrum, to a given cosmological model),
 and then
they are left to evolve according with  Newton's  laws and the
hydrodynamical equations.

\subsection{The Santa Barbara cluster test}
\label{stb}

The Santa Barbara cluster problem was proposed by \citet{Frenk:99}
to compare the results obtained from different codes. The
formation of a X-ray cluster in a Cold Dark Matter (CDM) universe
has been simulated using most of the hydrodynamic codes available
at that time, setting a standard of reference to test newly
proposed hydrodynamic codes.

The initial conditions of this test correspond to a $3\sigma$ peak
of the density field smoothed with a Gaussian filter of radius
$r_0=10$ Mpc according to the algorithm of \citet{Hoffm:91}. The
perturbation was centered on a periodic cubic region of side
$L=64$ Mpc. The cosmological scenario is a flat CDM universe with
$H_0=50$ km s$^{-1}$ Mpc$^{-1}$ for the Hubble constant;
$\sigma_8=0.9$ for the present-day linear rms mass fluctuation in
spherical top hat spheres of radius 16 Mpc; and $\Omega_b=0.1$ for
the baryon density (in units of the critical density). 64$^{3}$
dark matter and  64$^{3}$ baryon particles have been used with a
softening length of 20 kpc.

\begin{figure}
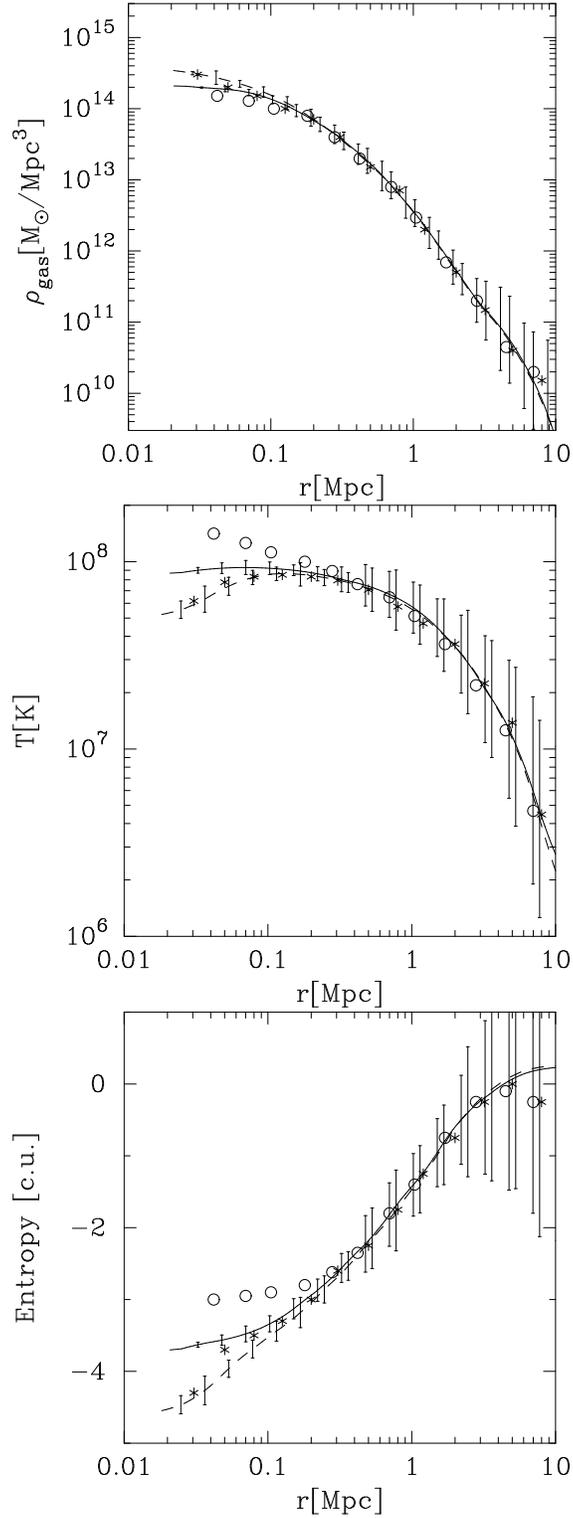

\centerline{\epsfig{figure=f4a.eps,width=0.45\textwidth}}

\centerline{\epsfig{figure=f4b.eps,width=0.45\textwidth}}

\centerline{\epsfig{figure=f4c.eps,width=0.45\textwidth}}

\caption{Density, temperature and entropy profiles in the Santa
Barbara cluster test. The stars represent the SPH results obtained
by Jenkins, while circles represent the results obtained by Bryan
\& Norman from an adaptive mesh refinement code. The lines
correspond to the results obtained from DEVA when the $\bn h$
terms are taken into account (solid line) and when these terms are
neglected (dashed line)} \label{stbfig}
\end{figure}

To test the influence of the $\bn h$ terms, two different
simulations were run with DEVA. In one of these simulations, the
$\bn h$ terms have been considered (SBGH) while in the other they
have been neglected (SBnoGH). In Figure \ref{stbfig}, the density,
temperature and entropy profiles of the cluster are plot. The
stars represent the results obtained \citep{Frenk:99} by Jenkins
from a high-resolution SPH simulation using a parallel version of
the Hydra code \citep{Pearc:97}, while the circles represent the
results obtained by Bryan \& Norman  from a high-resolution
adaptive mesh refinement shock-capturing code, SAMR,
\citep{Bryan:95,BryanN:95}. As previously remarked by
\citet{Frenk:99}, we see that the SPH and mesh results differ at
the central region. This figure also shows the results obtained
from our code both when the $\bn h$ terms are included (solid
line) and neglected (dashed line). Error bars correspond to the
standard deviation of the individual SPH data.

We see that, now, our results differ slightly depending on whether
the $\bn h$ terms have been included or not (the slope of the
density profile flattens more rapidly in SBGH than in SBnoGH; the
temperature profile is flat in SBGH and decreases within 100 kpc
in SBnoGH; the entropy profile is almost flat within 100 kpc in
SBGH and decreases continuously in SBnoGH). Moreover, when the
$\bn h$ terms are neglected, we obtain results that are similar to
those of previous SPH simulations, and very close to Jenkins'
results, obtained with a much higher resolution. When these terms
are taken into account, the results are intermediate between
previous SPH and grid results. This suggests that, at least in
part, the difference between the SPH and grid results could be due
to the non-physical entropy introduced by SPH codes. This
non-physical entropy is negative \citep{Alimi:02} and, therefore,
it produces objects with a smaller central temperature and a
higher central density.

\begin{figure}
\centerline{\epsfig{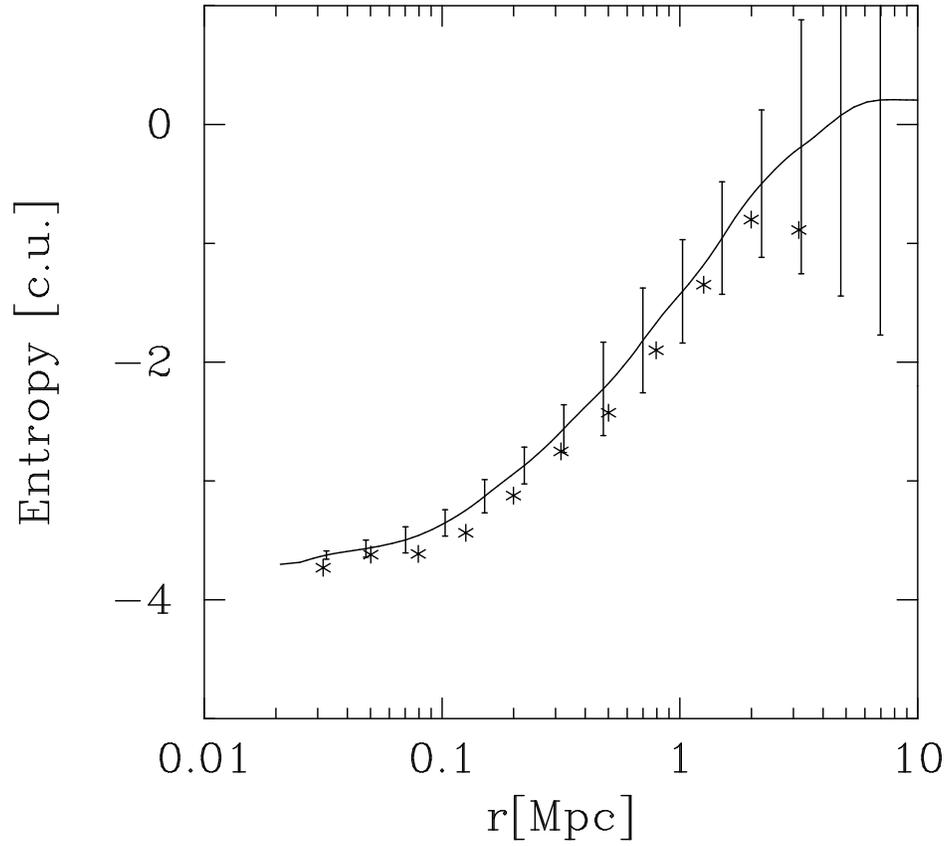}}

\caption{Comparison of the Santa Barbara entropy profiles obtained
from the entropy conserving versions of DEVA (solid line) and
GADGET (symbols)} \label{Comparison}
\end{figure}

Particular attention deserves the comparison of DEVA results with
those obtained from the entropy conserving SPH-Tree formulation by
Springel \& Hernquist (2002a), hereafter S-GADGET. In Figure
\ref{Comparison} we give a comparison of the Santa Barbara cluster
entropy profiles obtained in the SBGH run and S-GADGET, kindly
provided by Y. Ascasibar and G. Yepes \citep{Ascas:03}. Both have
been run with the same number of particles and gravitational
resolution. We see that the agreement is very good within the
error bars. So, both techniques compare very well in terms of
entropy conservation.

The differences found between SBGH and SBnoGH simulations can be
understood on the basis of the timescale for the $\bn h$ terms
(section 4.3). The time integral of $1/t_S$ (see Eq. \ref{ts}) for
this test is now larger than unity ($I=1.2$).

\subsection{Two Flat $\Lambda$CDM Cosmological Models}

We now report on several
 self-consistent simulations run in the context of
flat $\Lambda$CDM cosmological models with the aim of studying galaxy
assembly (see Table 2 for a summary).
We have considered two different  $\Lambda$CDM models  with
parameters whose values are  consistent with  their recent
determinations from observations: simulations $\Lambda$CDM1
($\Lambda$CDM2) have $\Omega_{\Lambda}=0.65$ (0.7), $\Omega_{\rm
baryon}=0.07$ (0.04), $\sigma_8=1.18$ (1.00) and $h=0.65$ (0.70)
\citep[][and references therein]{Lahav:02a,Nette:02,Sperg:03}. Both
simulations share the same seed for the Montecarlo realization of
the initial fluctuation field, so that each object formed in one
simulation has its counterpart in the other simulation. In each case,
we have
used $64^3$ DM particles and $64^3$ gas particles, in a periodic
box of 10~Mpc comoving side. The gravitational softening is
$\epsilon_{\rm g}=2.3$~kpc, and the minimum allowed smoothing
length, $h=\epsilon_{\rm g}/2$, as usual. For each cosmological
model,
two simulations have been run that are identical (they have
exactly the same initial conditions and the same values of the
cosmological parameters) except that in one case the $\bn h$ terms
have been included (simulations $\Lambda$CDM1-H and
$\Lambda$CDM2-H hereafter), while in the second case these terms
have not been taken into account (simulations  $\Lambda$CDM1-NoH
and $\Lambda$CDM2-NoH hereafter, see Table 2).
 Note that in these simulations
the cosmological volume is homogeneously sampled, in the sense
that no resampling multimass technique has been used to study GLO
assembly. This work can be considered as an extension of previous
works in standard CDM models
\citep{Domin:98,Saiz:01,Tisse:01,Tisse:02}, where cosmological
simulations had been run with a different code based on a
different numerical approach with fixed integration timesteps and
particle masses \citep{Tisse:97}. Relative to these previous
works, DEVA opens the possibility of considering the effects of
the $\bn h$ terms at scales of galaxies in self-consistent
simulations. Moreover, the simulations we report here represent an
improvement of the baryonic mass resolution by factors of $\simeq
10$ in the number of baryonic particles sampling a GLO of a given
total mass. Also, the time resolution allowed by the multistep
technique has been improved by a factor of $\simeq 30$ in the
denser areas of the box.

Cooling has been implemented as described in $\S$ 3.4, where the
cooling curve is that from \citet{Tucke:75} and \citet{Bond:84}
for an optically thin primordial mixture of H and He ($X=0.76$,
$Y=0.24$) in collisional equilibrium and in absence of any
significant background radiation field.

Concerning star formation, in this work we report on {\it direct}
results (i.e., no spectrophotometric) obtained with the simplest
implementation of star formation in the code: through a
parameterization, similar to those used by \citet{Katz:92} and
\citet{Tisse:97},  see  \citet{Alimi:02} for details, based on the
Jeans criterion for a collapsing region. Gas particles are turned
into stars according with an inefficient Schmidt-law-like
transformation rule \citep[see][]{Kenni:98,Silk:01},
\begin{equation}
\frac{d\rho_g}{dt}=-\frac{d\rho_{\ast}}{dt}=-\frac{c_{\ast}\rho_g}{t_g}\;,
\label{star-rate}
\end{equation}
where $c_{\ast}$ is a dimensionless star-formation efficiency
parameter, and $t_g$ is a characteristic time-scale chosen to be
equal to the maximum of the local gas-dynamical time
$t_{dyn}=(4\pi G\rho_g)^{-1/2}$, and the local cooling time,
$t_{cool} = u_i/\dot{u}_i$. Equation (\ref{star-rate}) implies
that the probability $p$ that a gas particle forms stars in a time
$\Delta t$ is
\begin{equation}
p=1-e^{-c_{\ast}\Delta t/t_g}\;.
\end{equation}

As usual, we compute $p$ at each  time step for all eligible
gas particles and draw random numbers to decide which particles
actually form stars.

Stellar feedback processes have not been explicitly considered,
but a tuning of the efficiency parameters
 can mimic these feedback effects.
Galaxy-like objects of different morphologies appear in the
simulation: disk-like objects (DLOs), early-type-like objects
(ETLOs) and irregular objects. DLOs contain gas in an extended
disk, and most stars in a  massive compact central concentration.
In simulations with lower $\epsilon_{\rm g}$ values (not reported
here),  stars form also in the disks, along arms. ETLOs are very
poor in gas and their stellar component have relaxed regular
ellipsoidal shapes. Irregulars have not defined shapes, and, in
most cases, they are the product of a recent merger or interaction
event. We note that GLOs formed in $\Lambda$CDM2-NoH and
$\Lambda$CDM2-H tend to be of later type than their
$\Lambda$CDM1-NoH or $\Lambda$CDM1-H counterparts, because of the
lower values the parameters $\Omega_{\rm baryon}$ and $\sigma_8$
take in $\Lambda$CDM2 simulations \citep{Domin:03}. First analyses
of GLO formation and evolution in simulations run with the DEVA
code are reported in \citet{Saiz:02a,Saiz:02b}; more details will
be given elsewhere (S\'aiz el al. 2003, in preparation). Here we
focus on different aspects related with DEVA performances.

\begin{figure}
\centerline{\epsfig{figure=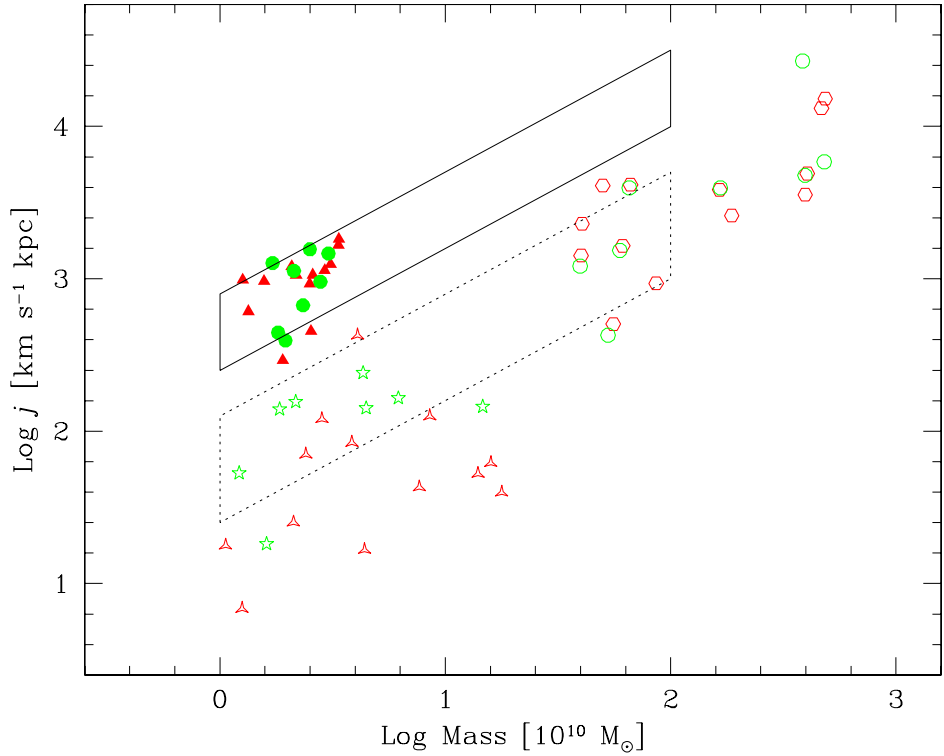,width=0.75\textwidth} }

\caption{Specific total angular momentum versus mass for
galaxy-like objects formed in $\Lambda$CDM2-NoH (triangles) and
$\Lambda$CDM2-H (pentagons) with $V_{2.2} \ge 150$ km s$^{-1}$ at
$z=0$. Open symbols: dark matter halos; filled symbols: the inner
83 per cent of the disc gas mass; starred symbols: central stellar
component (see text). The solid (dashed) box encloses the region
of the diagram corresponding to observed spirals (ellipticals)
}\label{JespMass}
\end{figure}

One important issue related to disc formation in hydrodynamical
simulations is specific angular momentum conservation at kpc
scales \citep{Somme:01,Stein:99,Domin:98,Saiz:01}. In Fig.
\ref{JespMass},   the specific total angular momentum at $z=0$ is
represented versus mass for DLOs identified in $\Lambda$CDM2-NoH
and $\Lambda$CDM2-H with $V_{\rm cir}(2.2 R_{\rm d}) \ge$ 150 km
s$^{-1}$ \citep[$R_{\rm d}$ is the disc scalelength,
see][]{Saiz:01}. The specific total angular momentum is plot for
dark haloes, $j_{\rm dm}$ (open symbols), for the inner 83 per
cent of the disc gas mass (i.e., the mass fraction enclosed by
$R_{\rm opt} \equiv 3.2R_{\rm d}$ in a purely exponential disc),
 $j_{\rm
g}$ (filled symbols), and for the stellar component of the DLOs in
our simulations, $j_{\rm s}$ (starred symbols). We see that,
except for the three more massive objects, $j_{\rm g}$ is of the
same order as $j_{\rm dm}$, so that these gas particles have
collapsed conserving, on average, their angular momentum.
Moreover, DLOs formed in our simulations are inside the box
defined by observed spiral discs in this plot \citep{Fall:83}. In
contrast, $j_{\rm s}$ is much smaller than either $j_{\rm dm}$ or
$j_{\rm g}$, meaning that the stellar component in the central
parts has formed out of gas that had lost an important amount of
its angular momentum in catastrophic events or that had never
acquired it. We note that this result is similar to those obtained
in simulations run with the code by \citet{Tisse:97}, even if the
codes are different, as explained above, and that the global
cosmological models are also different.

The smoothed estimate of the local gas density given by Eq. (2)
and the ensuing formulation of SPH equations, symmetrized to
ensure the reciprocity principle, has made  possible $j_{\rm g}$
conservation in an axisymmetric  potential. This is a delicate
crucial point in SPH codes. As stated in $\S$2.2,
its accurate implementation requires
that the individual smoothing lengths must be completely updated
at the beginning of each integration step,
what increases the  CPU time requirements. This complication cannot be
avoided, however, when an accurate $j_{\rm g}$ conservation is a
key point in  the physical processes under study. A different
approach to saving CPU time is then necessary. Using different
timesteps for each particle, useless computations for particles
that do not require high time resolution are avoided. This saves
considerable amounts of CPU time. Multistepping has allowed us to
run these simulations, that have a considerable dynamical range,
in a modest computing machine (a pentium IV 1.7GHz personal
computer, see above).

\begin{figure}
\centerline{\epsfig{figure=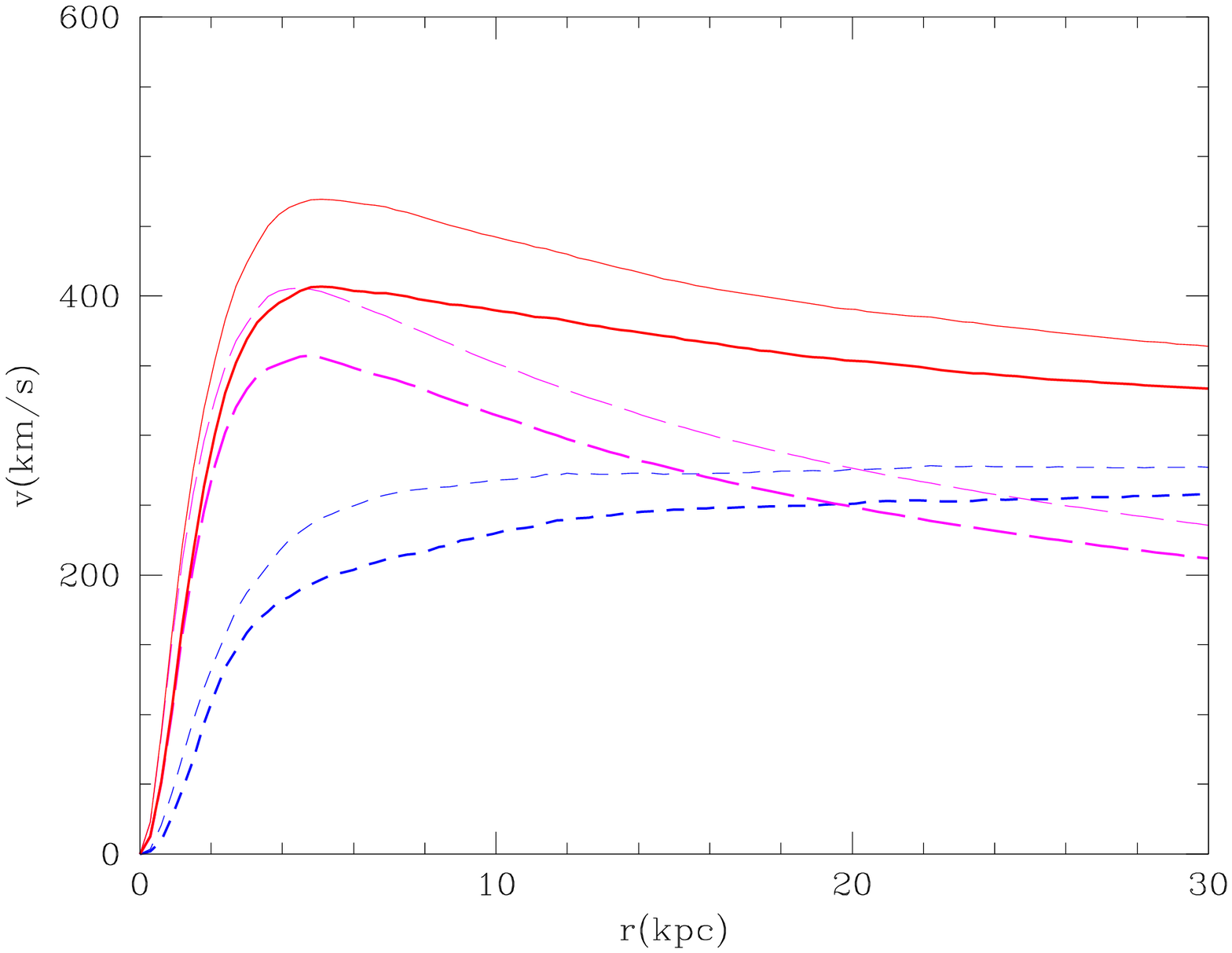,width=0.7\textwidth}}

\centerline{\epsfig{figure=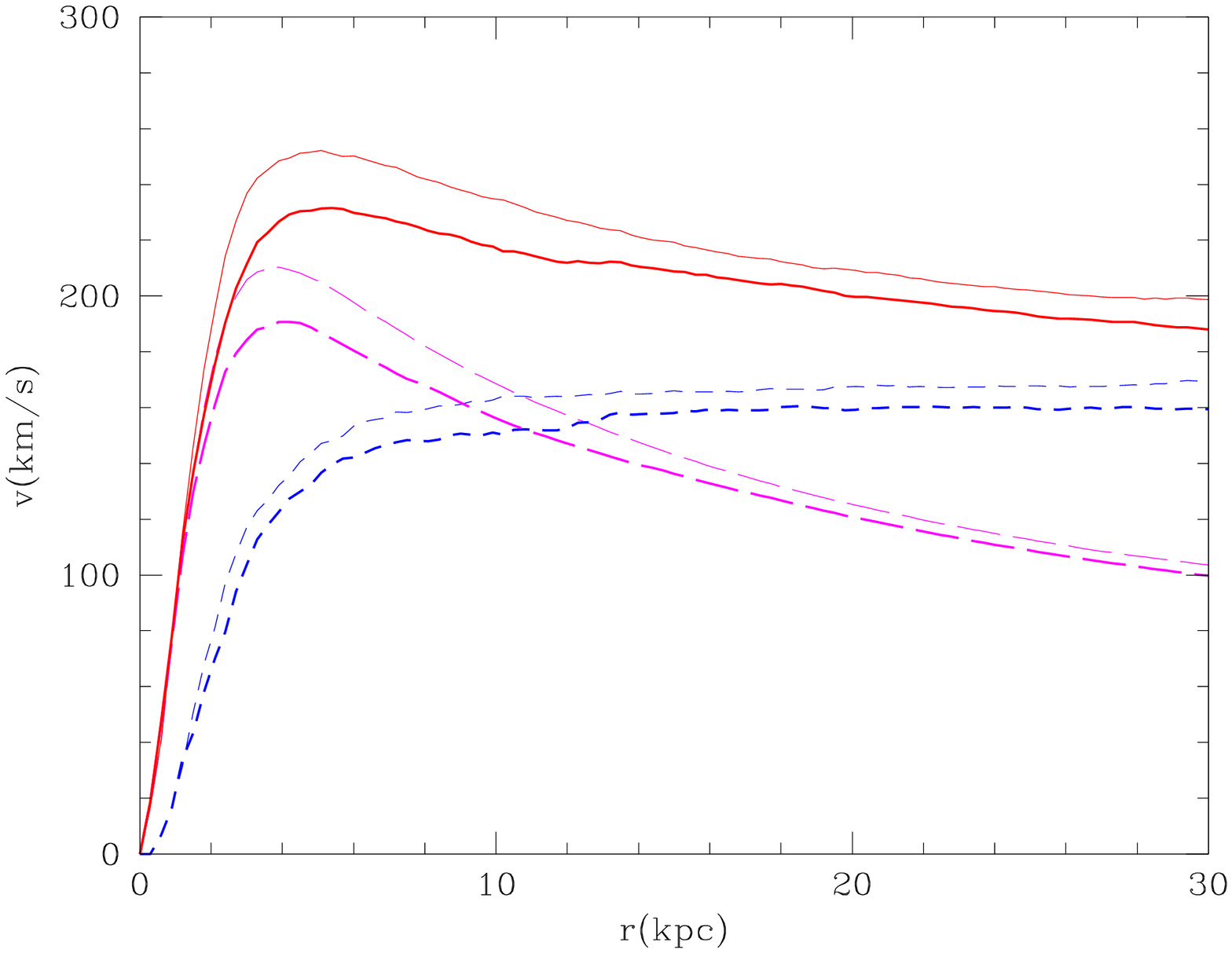,width=0.7\textwidth}}

\caption{ Circular velocity curves for two ETLOs (upper panel) and
two  DLOs (lower panel). ETLOs have formed in $\Lambda$CDM1-H
(thick lines) and $\Lambda$CDM1-NoH (thin lines); DLOs have formed
in $\Lambda$CDM2-H (thick lines) and $\Lambda$CDM2-NoH (thin
lines). Short-dashed and long-dashed lines stand for the dark
matter and the baryonic (both stellar and luminous) contributions
to the circular velocity (solid line) }\label{VelCir}
\end{figure}

Let us now turn to the effects of the $\bn h$ terms at kpc scales.
As stated in $\S$4.3, their general effect is to correct  the
spurious negative entropy introduced in their absence by SPH
codes, mainly at the central regions of collapsed objects. As a
consequence, when the $\bn h$ terms are taken into account,
dissipation by the gaseous component of a given GLO increases, so
that they are more disordered or dynamically hotter at their
central regions. Equilibrium is then attained with lower  central
baryon concentrations or densities, decreasing the amount of gas
infall. But this is not the unique effect of $\bn h$ terms on mass
distribution.  It is a well known effect that dark matter is
pulled in by baryons as they lose their energy and fall onto these
central volumes of collapsed configurations
\citep{Dalca:97,Tisse:98}. Decreasing the amount of gas infall
translates, as a consequence, into a decreasing of the amount of
dark matter at the GLO centers. As an illustration of this effect
on both the gaseous and dark components, in Figure \ref{VelCir}
(upper panel) we show the circular velocity curves for the most
massive ETLO formed in $\Lambda$CDM1-H (thick lines) and its
counterpart formed in $\Lambda$CDM1-NoH (thin lines). In the lower
panel, the corresponding circular velocity curves are represented
for a DLO formed in $\Lambda$CDM2-H (thick lines) and its
counterpart  in $\Lambda$CDM2-NoH (thin lines). In these Figures,
$r$ is the radial distance to the GLO
center of mass,
solid lines are the
circular velocities, $V^{2}_{\rm cir}(r) = V^{2}_{\rm dm}(r) +
V^{2}_{\rm bar}(r)$,
short-dashed lines and long-dashed lines stand for the dark matter
(dm)
and the baryonic (bar, both stellar and luminous) contributions,
respectively, given by:
\begin{equation}
V^2_{\rm i}(r) = \frac{G M_{\rm i}(<r) r^2}{(r^2 + \epsilon_{\rm
g}^2)^{3/2}}\;,
\end{equation}
with $i$ = bar and  dm.
 As can be clearly seen in these Figures, the
central distributions of both, dark matter and baryons, are
different and in any case the concentrations are lower when the
$\bn h$ terms are included. These central
concentrations are often
quantitatively estimated in literature through the $V_{\rm cir}^{\rm
peak}$ parameter \citep[the maximum or peak circular velocity,
see][]{Court:97,Saiz:01}. GLOs in $\Lambda$CDM1-H  or
$\Lambda$CDM2-H have lower $V_{\rm cir}^{\rm peak}$ values than
their $\Lambda$CDM1-NoH and $\Lambda$CDM2-NoH counterparts,
respectively. This can be seen in Figure \ref{VelCir} for an ETLO
and a DLO, but the behavior is general for any GLO. Other useful
parameter to quantitatively characterize circular velocity curves
is the logarithmic slope (LS), observationally defined for disc
rotation curves as the slope of the straight line that fits
$V_{\rm cir}(r)$ in log-log scale from $R_{2.2} = 2.2 R_{\rm d}$
up to the last measured point in the rotation curve \citep[][and
references therein; $R_{\rm d}$ is the disk scalelength]{Saiz:01}.
LSs are a measure of the GLO halo compactness at the scales of 10
- 30 kpcs. As illustrated  in Figure \ref{VelCir} for an ETLO and
a GLO, a  tendency for GLOs to be less compact also at these
scales when the $\bn h$ terms are taken into account has been
found in the simulations.

\begin{figure}
\centerline{\epsfig{figure=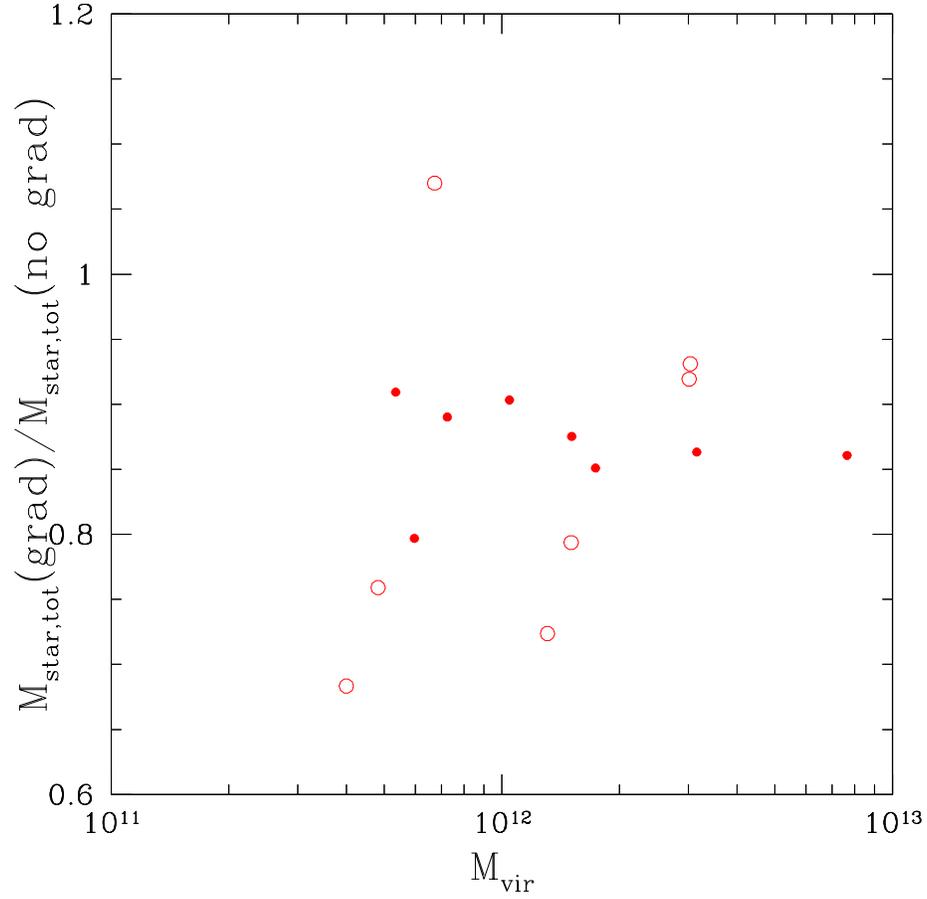,width=0.75\textwidth}}

\caption{ The total stellar mass inside the virial radii of ETLOs
formed in $\Lambda$CDM1-H (filled symbols) and $\Lambda$CDM2-H
(open symbols) in units of their $\Lambda$CDM1-NoH and
$\Lambda$CDM2-NoH counterparts (where the $\bn h$ terms have not
been taken into account), versus their virial masses }
\label{MassStar}
\end{figure}

A second effect of the $\bn h$ terms is related to the amounts of
stars formed in a given GLO at its formation and all along its
evolution until $z=0$. In these simulations, many stars form at
the shock fronts, where gas is compressed to very high densities.
Softer shocks mean less star formation with the same efficiency
parameters. To illustrate this effect, in Figure \ref{MassStar}
the ratios of the total stellar masses for the 8 (7) more massive
ETLOs produced in $\Lambda$CDM1-H ($\Lambda$CDM2-H) over  the
total stellar masses of their $\Lambda$CDM1-NoH
($\Lambda$CDM2-NoH) counterparts,
 are plot
versus their total virial masses, $M_{\rm vir}$.
 We see that, except in one
case, these ratios are smaller than the unity,
as expected.

\begin{figure}
\centerline{\epsfig{figure=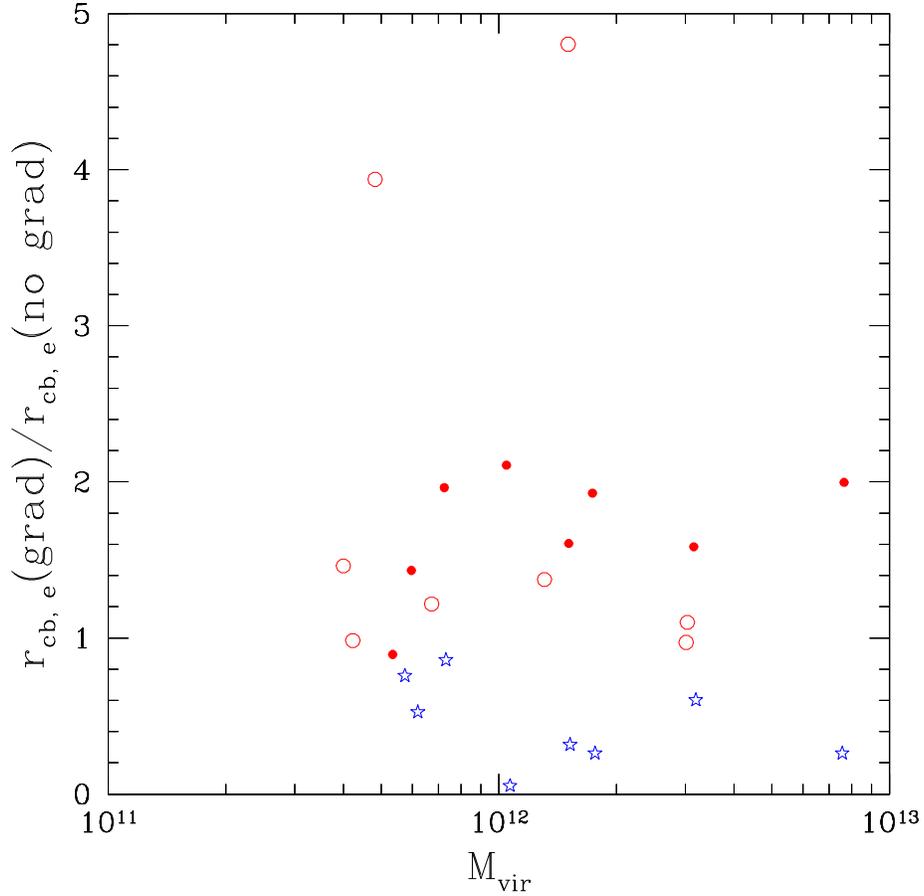,width=0.75\textwidth}}

\caption{ The 3D cold baryon effective radii of ETLOs formed in
$\Lambda$CDM1-H (filled circles) and $\Lambda$CDM2-H (open
circles) in units of their $\Lambda$CDM1-NoH and $\Lambda$CDM2-NoH
counterparts (where the  $\bn h$  terms have not been taken into
account), versus their virial masses. Starred symbols stand for 3D
cold baryon effective radii of ETLOs formed in $\Lambda$CDM1-MCool
in units of their $\Lambda$CDM1-NoH counterparts (see  Table and text) }
\label{reStar}
\end{figure}

Concerning sizes, for ETLOs we define the intrinsic or 3D cold
baryon effective radius, $r_{\rm cb, e}$, as the radius of the
sphere enclosing half the total  ETLO mass in cold baryons (i.e.,
cold gas or stars), ${M}_{\rm cb, tot}$. This is a measure of the
ETLO size at scales of the baryonic objects. In Figure
\ref{reStar} we plot $r_{\rm cb, e}$ for the $\Lambda$CDM1-H
($\Lambda$CDM2-H) ETLOs in units of  those of their
$\Lambda$CDM1-NoH ($\Lambda$CDM2-NoH) counterparts. As expected,
$\Lambda$CDM1-H  ($\Lambda$CDM2-H) ETLOS have larger baryonic
sizes than their $\Lambda$CDM1-NoH ($\Lambda$CDM2-NoH)
counterparts, except for one case.

\begin{figure}
\centerline{\epsfig{figure=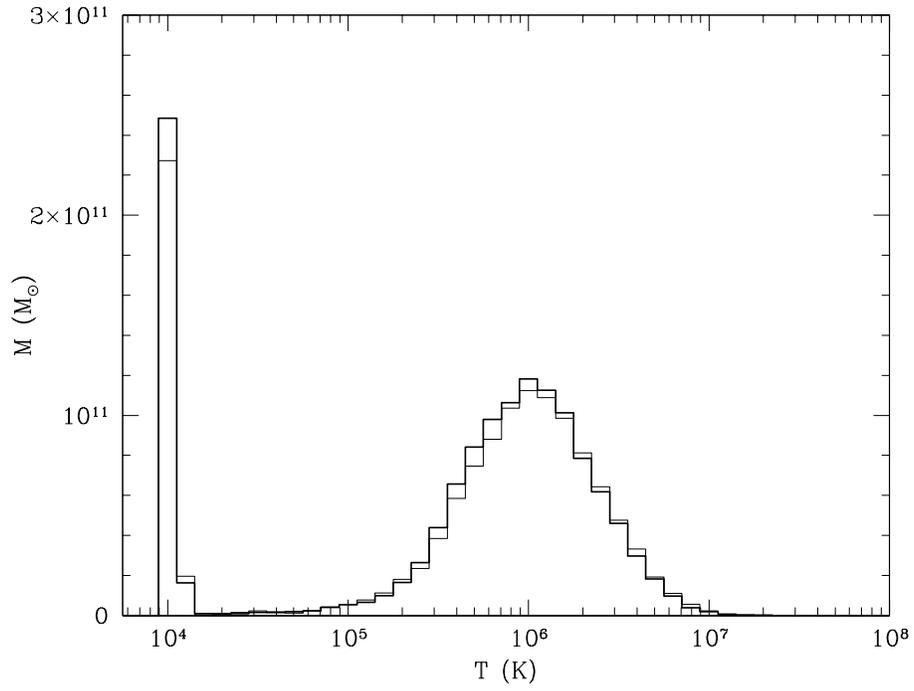,width=0.75\textwidth}}
\vspace{-2cm}

\centerline{\epsfig{figure=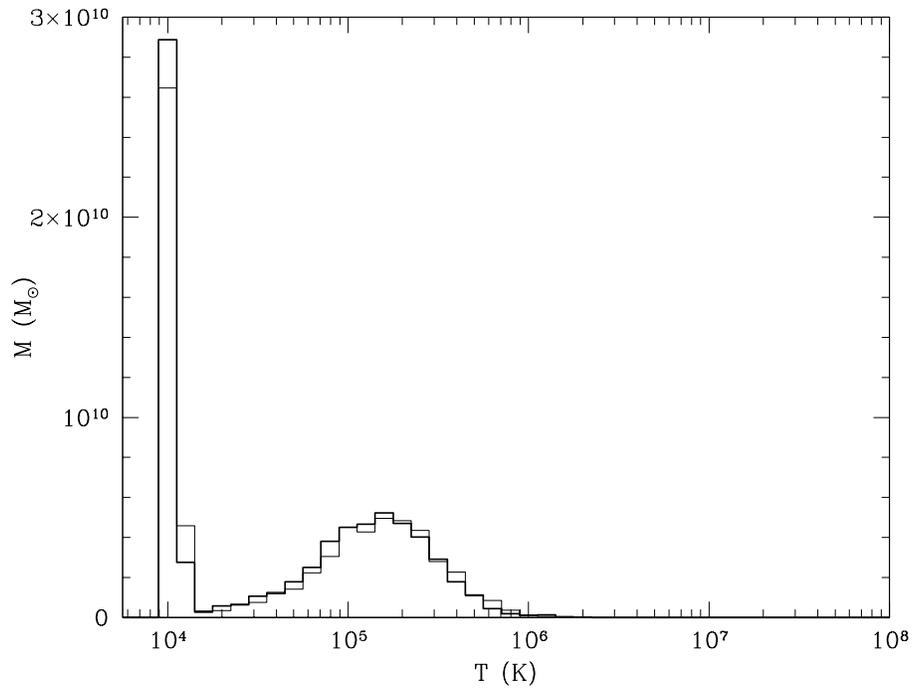,width=0.75\textwidth}}
\vspace{-2cm}

\caption{ Histograms for the temperature distribution of the gas
particles sampling  two of the  most massive ETLOs formed in
$\Lambda$CDM1-H (thick lines) and for their counterparts formed in
$\Lambda$CDM1-NoH (thin lines) } \label{TempHist}
\end{figure}

Finally, shocks heat the gas, producing an extended diffuse hot
component. The effect of the $\bn h$ terms in this case is to
lower the temperature of this diffuse gaseous halo component of
GLOs relative to the case when these terms are not considered.
Histograms for the temperature distribution of the gaseous
component of two of the  most massive GLOs in $\Lambda$CDM1-H
(thick lines) and in $\Lambda$CDM1-NoH (thin lines) are shown in
Figure \ref{TempHist}. To draw these histograms, all the gas
particles within a sphere of radius $r_{\rm lim}$, centered at the
GLO center of mass, have been considered  ($r_{\rm lim}$ is the
radius where the curve of integrated plasma emission luminosity
reaches its asymptotic value). As illustrated in the histograms
for these two GLOs, the temperature distributions of the gaseous
component for $\Lambda$CDM1-H and $\Lambda$CDM1-NoH objects do not
differ substantially. In both cases, the gas is close to biphasic,
but $\Lambda$CDM1-H objects are slightly colder within $r_{\rm
lim}$ than $\Lambda$CDM1-NoH objects. They are also slightly more
extended at $r_{\rm lim}$ scales, with the small excess of colder
particles placed at the outskirts of the gaseous configurations.

The differences found between $\Lambda$CDM1-H and
$\Lambda$CDM1-NoH or between $\Lambda$CDM2-H and $\Lambda$CDM2-NoH
simulations can be understood on the basis of the timescale for
the $\bn h$ terms (section 4.3). The time integral of $1/t_S$ (see
Eq. \ref{ts}) is now much larger than unity ($I\simeq80$),
indicating that the $\bn h$ terms cannot be neglected in this kind
of simulations.

As remarked in $\S$\ref{IncSPH}, cooling implementation in
multistep codes has to be handled with caution: cooling processes
need to be updated at each timestep for all particles. Otherwise,
gas particles involved in shocks suffer  from a spurious cooling
and non-physical cores of high density appear, giving rise to
extremely compact objects. As an example, in Figure \ref{reStar}
we plot the 3D cold baryon half-mass radii for a variant of
$\Lambda$CDM1-NoH, termed $\Lambda$CDM1-MCool (see Table 2
for details), where only active
particles at each timestep have been allowed to cool.
In Figure \ref{reStar} (starred symbols) we see that now  the
effective radii for the most massive objects are significantly
smaller than in $\Lambda$CDM1-NoH; the difference becomes less
important as the mass of the objects decreases and,
as a consequence, the fraction of
their constituent particles involved in shocks also decreases. Note
that, as the comparison of filled points and starred symbols in
Figure \ref{reStar} shows,  the combined effects of entropy
violation and spurious cooling can produce unphysically very small
objects, a factor of ten smaller, in some cases, than the values
found when these effects are circumvented (i.e., the
$\Lambda$CDM1-H simulation).

\section{Summary, Discussion and Conclusions}

We present DEVA, a multistep AP3M-like-SPH code designed to study
galaxy formation and evolution in connection with the global
cosmological model, that uses a formulation of SPH equations
ensuring both energy and entropy conservation.

Multistepping is introduced to save CPU time. In self-consistent
cosmological simulations multiple time scales appear, due to their
large dynamical ranges. To avoid  that particles in denser zones
slow down the simulation, and, at the same time, to get a properly
accurate integration algorithm, it is then advantageous to use
individual time steps. A comparison of the CPU time used in a
self-consistent cosmological simulation when it is run with a {\it
global} timestep or with a {\it multistep} scheme indicates that
in the second case  results at an equivalent level of accuracy are
produced $\sim$5  times faster.

When a multistep scheme is adopted, a delicate issue in the study
of galaxy assembly in a cosmological context is the {\it cooling}
implementation. In DEVA, as no  cooling timescale is taken into
consideration to fix the individual timestep for each particle,
cooling  must be calculated in a non-multistep way. Otherwise,
 particles involved in strong shocks would spuriously
cool and form dense objects characterized by
unphysically small baryon half-mass radii.

On writting DEVA, particular attention was paid that
conservation laws of  physics (energy, entropy, momentum) are
correctly implemented in the code, so that they hold at scales and
under physical conditions relevant for galaxy assembly in a
cosmological context. The usual formulations of SPH equations
focus on energy conservation and they violate, by construction,
entropy conservation.
 As a consequence, a negative entropy is numerically
produced in shocks that might (or might not) spuriously pollute
the results, depending on the problem one studies. Different
authors have addressed the issue of entropy violation in SPH codes
\citep{Nelso:93,Nelso:94,Hernq:93,Sprin:02a} and have given
different alternative formulations of its equations. We have
implemented a formulation that considers explicitly the effects of
the $\bn h$ terms in SPH equations \citep{Nelso:94}. By taking
advantage of the structure of the AP3M algorithm in the neighbor
search, the implementation of $\bn h$ terms in the code is simple
and noiseless.

To test the relevance of entropy violation at  shock
locations under   different physical
conditions, we have studied problems and run simulations (namely, the
adiabatic Santa Barbara cluster formation test, \citet{Frenk:99},
and fully self-consistent cosmological simulations to study galaxy
formation including  cooling an star formation) using identical
initial conditions and two different versions of DEVA, one that
takes into account the $\bn h$ terms in SPH equations and one that
does not take them into account. We show that entropy violation
has consequences  on the thermodynamical properties of the very
central regions of the Santa Barbara cluster and on the structure
at kpc scales  of galaxy-like objects (GLOs) formed in
simulations, but it does not have any appreciable consequence on
the standard  non-cosmological tests of hydrodynamical codes. To
understand the origins of these different behaviors, a criterion
is introduced that allows to elucidate when entropy violation is
expected to have  appreciable consequences on the results. In
standard tests, only moderate shocks and time scales are involved.

As a consequence of  the non-physical negative entropy numerically
produced when the $\bn h$ terms are neglected, both GLOs and the
cluster  present more concentrated baryon density profiles (either
star or gas). Concerning the Santa Barbara cluster test, when the
$\bn h$ terms are not included, we get results that are similar to
those of previous SPH simulations that focus on energy
conservation. However, when these terms are considered, the
results are intermediate between those SPH  and grid results. For
example, the temperature profile is decreasing within about 100
kpc of the cluster center when the simulation is run with standard
SPH codes (including DEVA without $\bn h$ terms), it increases
when it is run with grid codes and it is flat when  DEVA + $\bn h$
is used.
We would like to note that the accuracy figure of entropy
conservation obtained with DEVA + $\bn h$ and the
entropy-conserving SPH-Tree code by Springel \& Hernquist
 (S-GADGET, 2002a) compare quite satisfactorily,
as indicated by the good agreement between the
Santa Barbara cluster entropy profiles obtained with
both codes.

In cosmological simulations, negative entropy causes galaxy-like
objects (GLOs) to be  dynamically less hot  and gas infall
onto their central regions is
artificially increased, causing, also, an increase of the amount
of dark matter at the GLO centers.
These results
qualitatively agree with  those obtained with S-GADGET by
\citet{Sprin:02a} in their  simulation of a $\Lambda$CDM
cosmological model. No quantitative comparisons are
possible by the moment, because  a standard of comparison
for self-consistent cosmological simulations is
unfortunately not available.

An important result of this work is that the combined effects of
entropy violation and multi-step cooling implementation in
cosmological simulations can be particularly dramatic concerning
the concentration of mass distribution in the galaxy-like objects
they produce. For example,  their baryon half-mass
radii can be up to a factor of ten smaller than
half-mass radii of GLOs produced in entropy-conserving
non-multistep runs with DEVA.

Concerning momentum conservation, we have used a formulation of
SPH equations that is consistent with the smoothed estimate of the
local gas density (Eqs (1) and (3)). Equations are symmetrized to ensure
that the reciprocity principle holds (that is, if at a given time
the $j$th particle belongs to the neighbor list of the $i$th
particle, then it is mandatory that, at this same time, the $i$th
particle belongs to the neighbor list of the $j$th particle), so
that momentum and angular momentum are conserved. The
implementation of this principle in a SPH code increases
considerably the CPU time per integration step, because a double
loop on gas particles is necessary to evaluate smoothing lengths.
To test angular momentum conservation, we have measured the
specific angular momentum of discs formed in self-consistent
simulations. It has been found that conservation is good enough to
obtain simulated discs with observational counterparts, without
any need of previous heating, as already \citet{Saiz:01} have
shown \citep[see also][]{Gover:02}.

The use of a very high number of particles could ensure angular
momentum conservation in SPH-tree codes \citep{Gover:02}. In this
paper it has been shown that codes paying a particular attention
to the implementation of conservation laws  of physics at the
scales of interest can attain a good level of accuracy in
conservation laws with more limited resources.

\section*{Acknowledgments}

It is a pleasure to thank  A. Knebe,  M. Norman, V.
Quilis,  J. Silk, J. Sommers-Larsen, P. Tissera and G. Yepes for
useful information and discussions on the topics addressed in this
paper. Particular thanks are due to H.M.P. Couchman for making
public his AP3M code, on which the gravitational part of DEVA is
based, and to Y. Ascasibar and  G. Yepes for allowing us to use
their results on the Santa Barbara cluster test.

 This project was partially supported by the MCyT (Spain)
through grant AYA-0973  from the Programa Nacional de
Astronom\'{\i}a y Astrof\'{\i}sica. We also thank the Iberdrola
Foundation for financial support, and the Centro de Computaci\'on
Cient\'{\i}fica (UAM, Spain) for computing facilities.

\bibliographystyle{mn}

\newpage

\begin{table}
\begin{center}
{\sc TABLE 1}

{\sc Entropy and Energy Conservation}

\begin{tabular}{cllll} \hline\hline
Run & \multicolumn{1}{c}{$\bn h$ terms} &\multicolumn{1}{c}{$N$} &
\multicolumn{1}{c}{$\Delta E$} & \multicolumn{1}{c}{$\Delta S$}\\
\hline
HTest1 & Excluded  & 1024 & 0.01\% & 5.6\%  \\
HTest2 & Included  & 1024 & 0.02\% & 0.02\% \\
HTest3 & Excluded  & 2048 & 0.01\% & 5.2\%  \\
HTest4 & Included  & 2048 & 0.01\% & 0.01\% \\
HTest5 & Excluded  & 4096 & 0.02\% & 5.5\%   \\
HTest6 & Included  & 4096 & 0.01\% & 0.02\% \\
\hline
\end{tabular}
\end{center}
\end{table}

\newpage

\begin{table}
\begin{center}
{\sc TABLE 2}

{\sc Summary of $\Lambda$CDM Simulations}\vspace{0.3cm}

\begin{tabular}{lrrrrcc} \hline\hline
\multicolumn{1}{c}{Run} &\multicolumn{1}{c}{$\Omega_{\Lambda}$} &
\multicolumn{1}{c}{$\Omega_{B}$} & \multicolumn{1}{c}{$\sigma_8$}
& \multicolumn{1}{c}{$h$} & \multicolumn{1}{c}{$\bn h$ terms}
& \multicolumn{1}{c}{Cooling}\\
\hline
$\Lambda$CDM1-H     & 0.65  & 0.07 & 1.18 & 0.65 & Yes & Non-multistep  \\
$\Lambda$CDM1-NoH   & 0.65  & 0.07 & 1.18 & 0.65 & No  & Non-multistep \\
$\Lambda$CDM1-MCool & 0.65  & 0.07 & 1.18 & 0.65 & No  & Multistep  \\
$\Lambda$CDM2-H     & 0.70  & 0.04 & 1.00 & 0.70 & Yes & Non-multistep\\
$\Lambda$CDM2-NoH   & 0.70  & 0.04 & 1.00 & 0.70 & No  & Non-multistep   \\
\hline
\end{tabular}
\end{center}
\end{table}

\clearpage

\label{lastpage}
\end{document}